\documentclass[aps,twocolumn,superscriptaddress,groupedaddress]{revtex4}  
\usepackage{graphicx}  
\usepackage{dcolumn}   
\usepackage{bm}        
\usepackage{amssymb}   
\usepackage{amsmath}
\usepackage{appendix}

\RequirePackage{filecontents}

\usepackage{graphicx,subcaption}
\usepackage{dcolumn}
\usepackage{bm}
\usepackage{subcaption}

\begin{document}


\title{ Chaotic Dynamics in a Galactic Multipolar Halo with a Compact Primary }

\author{Yeasin Ali}
\altaffiliation[Also at ]{Dept. of Physics, St. Xavier's College, 30, Mother Teresa Sarani, Kolkata-700016, India} 
\email{ali.yeasin3.0@gmail.com}

\author{Suparna Roychowdhury}

\affiliation{
	Dept. of Physics, St. Xavier's College, 
	30, Mother Teresa Sarani, 
	Kolkata-700016, India
}


\begin{abstract}

Observational evidence strongly supports the existence of a Super Massive Black Hole (SMBH) at the Galactic center, surrounded by dense stellar clusters. Modeling galactic centers with intricate structures like shells and rings pose challenges, prompting the use of simplified models such as a spherical monopole potential with a multipolar halo mass distribution. This approach, employing a multipolar expansion model, provides versatility for numerical analyses, revealing the complex dynamics of stars in this region. Pseudo-Potentials like Paczynsky-Wiita and Artemova-Bjornsson-Novikov are utilized to simulate the impacts of strong gravity from non-rotating and rotating compact objects respectively, elucidating their influence on stellar dynamics. Chaos naturally arises due to non-central forces, visualized using the Poincaré section technique. Of particular importance is the utilization of the Smaller Alignment Index (SALI), a powerful nonlinear dynamical tool, which categorizes particle orbits as escaping, regular, sticky, or chaotic. We exhaustively examine all combinations of multipolar moments up to the octupolar term along with spin using this tool, which had not been studied earlier. SALI provides a straightforward yet efficient method for assessing the interplay between the system's different multipolar moments, their combinations, and spin. Thus, our findings offer insights into the dynamics of compact objects enshrouded in a halo mass distribution and lay the groundwork for understanding complex astrophysical systems in galactic centers.

\end{abstract}
\maketitle

\section{Introduction}
The Galactic center serves as a distinctive arena for investigating fundamental physical processes at the cores of galaxies. Observational evidence strongly support the presence of a Super Massive Black Hole (SMBH) \cite{BHKormendy1995}, \cite{AGNBeckmann2012}, \cite{BHKormendy2004}, \cite{Hawthorn2016GalaxyReviewArt} and dense stellar clusters \cite{starClusterBrok2014}, \cite{starClusterTurner_2012}, \cite{starClusterGeorgiev2014}, \cite{starClusterGeorgiev2016}, \cite{Genzel2010Galaxy}, \cite{Hawthorn2016GalaxyReviewArt} as predominant constituents in the central regions or bulges of galaxies. Bulges exhibit various shapes such as classical bulges as well as boxy-peanut bulges \cite{Hawthorn2016GalaxyReviewArt}. The compact stellar clusters found in galactic centers exhibit masses ranging from  $10^5$ to $10^8 M_\odot$ \cite{Hawthorn2016GalaxyReviewArt}, \cite{Genzel2010Galaxy}. In almost all instances, the SMBHs are typically enveloped by a hollow spherical halo of matter and extensive accretion disks \cite{Genzel2010Galaxy}, \cite{Hawthorn2016GalaxyReviewArt}. A myriad of intricate processes is associated with SMBHs and their immediate surroundings, making the study of star dynamics in the proximity of the galactic core a focal point of attention in recent decades.

Various models, such as the Hernquist model \cite{Hernquist1990GalacticPot} and Navarro, Frenk,  White (NFW) \cite{NFW1996GalacticPot} model, have been widely employed to describe the galactic potential and investigate stellar orbits in galaxies \cite{Lawrence2005GalacticPot}, \cite{binney2008galactic}. Modeling galactic centers, which often feature intricate structures like shells, rings and dense stellar clusters surrounding a compact object \cite{BHKormendy1995}, \cite{AGNBeckmann2012}, \cite{BHKormendy2004}, \cite{Hawthorn2016GalaxyReviewArt}, \cite{Malin1983GalaxyShell}, \cite{Bournaud2003GalaxyRing}, \cite{Reshetnikov1997GalaxyRing}, \cite{Sackett1990GalaxyRingHalo}, presents a significant challenge in astrophysics. The considerable difference in mass between a star and a SMBH permits a simplification of the problem. On intermediate distance and time scales, stars can be treated as test masses, ensuring that gravitational interactions between stars are negligible \cite{Alexander2017StellarDynamicsGalacticCenter}. Thus to address this complexity, simplified models commonly employ, such as a spherically symmetric monopole potential surrounded by a multipolar halo mass distribution, restricted to the core and the halo \cite{binney2008galactic}. By utilizing this multipolar expansion potential model, researchers have gained access to the versatility necessary for conducting analytical as well as computational modeling of the complex dynamics of different components within galactic cores and their surrounding environments. Typically such models employ the technique of using stars close to the galactic core as test particles moving under the influence of this multipolar potential in the center as a justified scheme in the galactic center studying the evolution of stellar orbits for long time scales. A comprehensive review of such work can be found in literature \cite{Vogt&Letelier2005Dipole&OctupoleMomentGalaxy}, \cite{FouvryDehnenScott2022MultipolePotGalaxy}, \cite{Meza2005MultipolePotentialGalaxy}, \cite{Meiron2014GPUMultipolePotGalaxy}, \cite{McGlynn1983MultipolePotGalaxy}. Despite its simplicity, this modeling technique offers valuable insights into the intricate dynamics of galactic cores.



Given the presence of an SMBH in the galactic center, a general relativistic approach can also be employed to understand particle dynamics around this compact object. However, due to the highly nonlinear nature of Einstein's field equations, simulating particle dynamics around such compact objects is computationally very intensive. To simplify calculations and computational modeling, researchers have proposed various  pseudo-potential techniques, such as those introduced by Paczynski $\&$ Wiita (1980) \cite{Pwitta1980} and Nowak $\&$ Wagoner (1991) \cite{Nowak1991SchwarchildPseudoPotential} to precisely capture the features of the space-time around non-rotating compact objects resembling Schwarzschild-like black holes. The features of the space-time surrounding rotating Kerr-like compact objects have been well emulated by pseudo-potentials proposed by Artemova et al. (1996) \cite{Artemova1996ModifiedNP}, Chakrabarti and Khanna (1992) \cite{Chakrabarti1992PseuPot}, Chakrabarti and Mondal (2006) \cite{Chakrabarti2006PseuPot}, Ghosh and Mukhopadhyay (2007) \cite{Ghosh2007PseuPot}, Ghosh et al. (2014) \cite{ghosh2014PseuPot}, Karas and Abramowicz (2014) \cite{Karas2014PseuPot}, Mukhopadhyay (2002) \cite{Mukhopadhyay2002PseuPot}, Semerák and Karas (1999) \cite{Semerak1999PseuPot} to faithfully. Among these alternatives, we have specifically selected the Paczynsky-Wiita (PW) potential  \cite{Pwitta1980} and Artemova–Björnsson–Novikov (ABN) \cite{Artemova1996ModifiedNP}  for our study with Schwarzschild-like and Kerr-like compact objects, respectively. The ABN potential allows for an exploration of the complete spectrum of behaviour from a rotating Kerr-like object to a Newtonian star and also a non rotating compact object in a single unified framework.


In this present work, our focus primarily lies on a two-body system: one comprising of a rotating or non-rotating compact object surrounded by an asymmetric multipolar halo mass distribution, and the other being a test particle with unit mass. To represent the influence of a black hole with this asymmetric multipolar halo mass distribution, we utilize a multipole expansion potential \cite{binney2008galactic}, \cite{vieira1997IntegrabilityHalo}, \cite{Letelier1997ChaosRotBHDipolerHaloGR}, \cite{sankha_Nag2017}, \cite{Ying2012}. We specifically employ a pseudo-potential to model the influence of a central compact object, which is represented by the monopole term in a multipolar expansion. This pseudo-potential is combined with dipolar, quadrupolar, and octupolar terms to account for the asymmetric mass distribution in the halo. Our objective is to comprehend the interactions between individual terms and combinations of various multipolar terms, such as the dipole, the quadrupole and the octupole with spin, and how they influence the dynamics of the of the test particle in our system.

In the absence of perturbative effects from the halo, the central black hole potential alone does not allow for chaotic motion. However, the introduction of even a minor non-central force renders the system non-integrable, as demonstrated by some authors \cite{Goldstein2002}, \cite{Contopoulos2002Book}. Therefore, when examining the dynamics of a test particle's orbit within a halo surrounding a compact object, the emergence of chaotic phenomena becomes a natural outcome. Considering that multipole moments reflect the inherent structure of a source, numerous studies within the realm of general relativity have explored their impact on the trajectory of test particles moving along geodesics (e.g., \cite{Gair2008ObservationalProperty..BumpySpacetime}, \cite{RamosCaro2011MotionAroundMonopole..ringSystem}, \cite{Hou&Zhaoyi2018BlackHoleShadowDipolarHalo}, \cite{Letelier1997ChaosRotBHDipolerHaloGR}, \cite{Gueron2002}, \cite{Letelier2003}, \cite{vieira1996Henon_HeilDipHaloGR}, \cite{Janiuk2011GalaxyAccretionGR}, \cite{Vieira1999CoreShellGR}, \cite{Han2008_ChaosDynamics..KerrSpacetime}).  Vieira \& Letelier (1996) \cite{vieira1996Henon_HeilDipHaloGR} have also investigated the effects of quadrupolar and octupolar moments in a system comprising a halo and a black hole in general relativistic formalism. They found that while the quadrupolar term alone does not induce chaos, the octupolar term serves as a significant catalyst. They  extended their earlier work to include Newtonian and relativistic core-shell models in \cite{Vieira1999CoreShellGR}, revealing significant effects on the orbital of stars inside galaxies. It was noted that in oblate shell models, the breakdown of symmetry increases chaos whereas in prolate shell models, it reduces chaos allowing for stable periodic orbits to emerge. Subsequent studies by de Moura \& Letelier (2000) \cite{deMoura2000chaosFractals..Halos} examined particle scattering behavior in core-shell gravitational models, finding no observable chaos with oblate halos. Guéron \& Letelier (2002) \cite{GueronLetelier2002GeodesicChaosQuadrupolar} and Dubeibe et al. (2007) \cite{Dubeibe2007ChaoticDynamics..NonisotropicStresses} independently studied geodesic motion around astrophysical entities with non-isotropic stresses, identifying potential chaotic trajectories in both oblate and prolate deformations, contingent upon specific quadrupolar deformities. Dubeibe et al. (2021) \cite{Dubeibe2021Quadrupole+Octupole} conducted a comparative analysis between Newtonian and relativistic frameworks for a composite system consisting of a halo and a black hole, examining the influence of quadrupolar and octupolar moments on the system's dynamics. Pseudo-potential schemes were also used to investigate chaotic dynamics of particles around both rotating and non-rotating black holes with halos. Guéron \& Letelier (2001) \cite{Gueron2001} concluded that general relativistic orbits are more stable as compared to Newtonian and special relativistic orbits. They also observed that the special relativistic orbits are least stable for a non-rotating black hole with a halo system. Investigations into compact rotating black holes surrounded by quadrupolar halos were conducted by Ying et al. (2012) \cite{Ying2012} to demonstrate that chaos is predominant for small spin values in systems comprising of halos with only a quadrupolar term (Q). Rotational parameters weaken chaos but counter-rotating systems, specially in prolete halos, are seen to be more unstable. Nag et al. (2017) \cite{sankha_Nag2017} investigated the correlation between the spin parameter and chaos in a system comprising of a central rotating black hole surrounded by a dipolar halo. They also observed a negative correlation between the black hole's spin parameter and the level of chaos exhibited by test particles within the dipolar halo. Additionally, they found that the equations of motion, without special relativistic corrections, closely approximate the behavior governed by the corrected special relativistic equations, albeit with a slightly higher value of angular momentum.

An essential element of chaos in galactic dynamics is illustrated by the phenomenon known as `stickiness' or weak chaos. In this scenario, orbits might not densely populate phase space as they do in strongly chaotic situations. Sticky orbits exhibit intermediate behavior between regular and chaotic orbits in dynamical systems, particularly in galactic dynamics. Unlike regular orbits, which have integrals of motion (including a third integral) and show predictable, quasi-periodic behavior, sticky orbits initially mimic this regularity, adhering closely to invariant tori (KAM tori) for extended periods. However, they eventually transition to chaotic behavior, lacking a third integral of motion and displaying unpredictability. This stickiness arises near the boundaries between regular and chaotic regions in mixed phase space systems \cite{Contopoulos2008_StickyOrbits}. Several approaches exist to
detect and quantify chaos, such as the Lyapunov Characteristics Exponents, the Lyapunov Characteristics Number, the Lyapunov indicator, Smaller Alignment Index (SALI) and many others. Recent literature extensively evaluates and discusses the differences and efficacy of these indicators \cite{Maffione2011ChaoticIndicatorComparison}, \cite{Maffione2013ChaoticIndicatorComparison}, \cite{Carpintero2014LPVIcodeAP}, \cite{Skokos2016BookChaos&Predictability}. One of the oldest and widely used tool to quantify chaos in a system is the Poincaré section (PS). In this study, we have employed the Poincaré section technique and SALI for quantafying the chaotic nature of the dynamics of test particle in this system. Several author (\cite{vieira1997IntegrabilityHalo}, \cite{Vieira1999CoreShellGR}, \cite{vieira1996Henon_HeilDipHaloGR}, \cite{Letelier1997ChaosRotBHDipolerHaloGR}, \cite{Letelier2011BHQpoleNewtonian}, \cite{sankha_Nag2017}, \cite{Ying2012}, \cite{saikat_das_SRC2023}) employ the PS method to discern chaos in a system containing a central BH and halo. The  Smaller Alignment Index (SALI) method, a relatively recent approach for identifying chaos, has been developed and applied across diverse dynamical systems \cite{Skokos2001}, \cite{Skokos2004}, \cite{Skokos2014SALI&GALIChaosDetection}, \cite{Skokos2016BookChaos&Predictability}. The SALI method is highly efficient and rapid in discerning chaos and order within a system, as has been observed by some earlier authors \cite{Maffione2011ChaoticIndicatorComparison}, \cite{Maffione2013ChaoticIndicatorComparison}, \cite{Carpintero2014LPVIcodeAP}, \cite{Skokos2016BookChaos&Predictability}. Zotos et al. have also extensively employed this technique in investigating various astrophysical systems (\cite{Zotos2022ChaosOrderBarredGalaxy}, \cite{Zotos2018OrbitalClassification}, \cite{Zotos2018SunJupitar}, \cite{zotos2014InterplaybetweenDark..EllipticalGalaxies}, \cite{zotos2016ClassificationOrbitsAxisymmetricDiskyGalaxies}, \cite{zotos2016EscapeDynamicsBinarySystem}, \cite{zotoz2017UnravellingEscapeDyna..starCluster}, \cite{Zotos2016FugitiveStar}, \cite{zotos2014CalssifingOrbitsinGalxy..DMatterHaloComponenet}, \cite{zotos2021QuantitativeOrbitClassification..SatelliteAroundJupiter} and \cite{Zotos2019NumericalInvestigationPlaneCircularPluto}) refences there in.

Current observations of stars and accreting matter in the centers of galaxies does indicate that their dynamics is often nonlinear and chaotic due to the influence of the massive central compact object and the distributed halo in the galactic center. Hypervelocity stars (HVSs) are propelled by the Milky Way’s central massive black hole (MBH) at velocities surpassing the Galactic escape velocity. These HVSs have been detected in the halo of the Milky Way \cite{Brown_2015HypervelocityStar}. It has been observed that stars from the globular cluster are being expelled. This phenomenon is believed to be primarily due to the gravitational interactions within the cluster and the tidal forces exerted by the supermassive black hole at the center of the Milky Way \cite{Roman_2023_StellarStream}, \cite{Weatherford2023StelarEscapeMechanism}. 
Observations of X-ray binaries, in which a stellar-mass black hole accretes matter from a companion star, exhibit quasi-periodic oscillations (QPOs) in the X-ray flux. These QPOs are believed to be associated with the dynamics of particles within the inner regions of the accretion disk \cite{Ingram2019QPO}, \cite{Ren2023QPO}. 

In this paper, we have organized our studies as follows: in section 2, we present the mathematical formulation of the problem; in section 3, we discuss how chaos is interlinked with the multipolar moments and spin of the BH using the PS for both Newtonian and Special Relativistic formalisms; in section 4, we have extensively used SALI to analyze the dynamics of the system, examining how multipolar moments and spin affect the dynamics and categorizing, the orbits of the particle into four categories, namely escaping, regular, sticky, and chaotic and finally, in Section 5, we provide our results and discussion.

\section{Mathematical Formulation}
The mass distribution of galaxies in its core with a central compact object can  be described well with multipole expansion scheme, as shown by several authors earlier \cite{binney2008galactic}. We have used this scheme up to the 3rd order (octupolar term) of the expansion to generate the potential of the central galactic bulge along with a compact object for our model. The monopole and the quadrupole ($Q$) terms represent the spherically symmetric SMBH and a halo shell (oblate, $Q>0$ in this work) around the SMBH respectively. The dipolar ($D$) and the octupolar ($O$) terms are introduced to account for the off-equatorial and other asymmetric effect of the halo mass distributions. We have replaced the spherically symmetric monopole term of the multipolar expansion with  Paczynsky-Wiita (PW) and Artemova–Björnsson–Novikov (ABN) pseudo-potential to study the dynamics of a particle around Schwarzschild and Kerr-type compact objects. Additionally, we have extended our investigation to include the simple Newtonian potential also for a comprehensive comparison and understanding of the complex dynamics in this scenario.

The free fall acceleration on a Kerr object with rotating parameter $a$ was proposed \cite{Artemova1996ModifiedNP} as, 
\begin{align} \label{Eqn:artimovaForce}
    F&=-\frac{1}{r^{2-\beta}(r-r_1)^{\beta}}
\end{align}
where, \( r_1 = 1 + \sqrt{1 - a^2} \) denote the radial position of the event horizon in the equatorial plane, expressed in units of \( {GM_{BH}}/{c^2} \), where \( G \) is the universal gravitational constant, \( M_{BH} \) is the mass of the black hole, \( c \) is the speed of light in a vacuum, and \( a \) is the Kerr parameter. We adopt the convention \( G = c = M_{BH} = 1 \) \cite{sankha_Nag2017}. With  
\begin{align*}
\beta &= \frac{r_{in}}{r_1}-1, \\
r_{in} &=3+Z_2-[(3-Z_1)(3+Z_1+2Z_2)]^{\frac{1}{2}}, \\
Z_1 &=1+(1-a^2)^{\frac{1}{3}}[(1+a)^{\frac{1}{3}}+(1-a)^{\frac{1}{3}}], \\ 
Z_2 &= (3a^2+{Z^{2}_1})^{\frac{1}{2}}
\end{align*}
The versatility of the ABN potential allows us to encompass the PW potential when the spin parameter `$a$' is set to 0, thereby reproducing the characteristics of a Schwarzschild like black hole. Furthermore, by setting the spin parameter `$a$' to 1, the ABN potential transforms into a simple Newtonian potential. This allows us to explore the entire spectrum of behaviour, starting from a non-rotating compact object ($a\;=\;0$) to a simple Newtonian object ($a=1$), with a rotating compact object being represented by $0<a<1$. 

We can calculate the ABN potential for this free fall acceleration by simply integrating it and setting the integration constant for asymptotically vanishing potential at infinity, $\phi=0$ for $r\to \infty$.
\begin{align}\label{Eqn:artimova_pot}
\Phi_{ABN}(r) = -\frac{1}{r_1(\beta - 1)}\left[\frac{r^{\beta-1}}{(r-r_1)^{\beta-1}}-1\right],
\end{align}
In our case, due to the presence of azimuthal symmetry, we can rewrite this potential in cylindrical polar coordinates $(\rho,\phi,z)$ as
\begin{equation}\label{Eqn:artimova_potCyl}
    \Phi_{ABN}(\rho,\phi,z) = -\frac{1}{r_1(\beta - 1)}\left[\frac{(\rho^2+z^2)^{\frac{\beta-1}{2}}}{(\sqrt{\rho^2+z^2}-r_1)^{\beta-1}}-1\right]
\end{equation}
The above equation Eq.~(\ref{Eqn:artimova_potCyl}) does not contain any $\phi$-dependence because of its azimuthal symmetry. \\
For non rotating cases, the rotating parameter $a$ becomes zero, implying that the ABN potential, given in Eq.~(\ref{Eqn:artimova_potCyl}), takes the form of the Paczynski-Witta as described below:
\begin{equation}\label{Eqn:PWittaPot}
    \Phi_{PW}(\rho,\phi,z) = -\frac{1}{\sqrt{\rho^2+z^2}-2}
\end{equation}
Similarly, when the rotating parameter $a=1$, the Artemova–Björnsson–Novikov (ABN) potential in Eq.~(\ref{Eqn:artimova_potCyl}) takes the form of a simple Newtonian potential given by, 
\begin{equation}\label{Eqn:NewtonianPot}
    \Phi_{N}(\rho,\phi,z) = -\frac{1}{\sqrt{\rho^2+z^2}}
\end{equation}
As previously mentioned, if the halo's mass distribution around the black hole is axially symmetric about the spinning axis and approximately spheroidal but slightly displaced from the equatorial plane with other asymmetries, the gravitational potential resulting from these asymmetric halo mass distribution along with a central compact object can be expressed as a summation of the dipolar, quadrupolar, and the octupolar terms superimposed on the central monopole. Consequently, the total gravitational potential acting on a test particle within the halo, can be formulated as,
\begin{equation} \label{Eq:potentialPhig}
    \Phi_g = \Phi (\rho,\phi,z)+Dz+\frac{Q}{2}(2z^2-{\rho}^2)+\frac{O}{2}(2z^3-3z{\rho}^2)
\end{equation}
where  \(\Phi(\rho,\phi,z)\) represents the potential associated with the centrally located compact object and $D, Q, O$ represent the dipole, quadrupole and octupole strengths respectively, as has been shown by several earlier works \cite{binney2008galactic}, \cite{vieira1997IntegrabilityHalo}, \cite{Letelier1997ChaosRotBHDipolerHaloGR}, \cite{Gueron2001}. To suit our specific requirements, we substitute \(\Phi(\rho,\phi,z)\) with \(\Phi_{ABN}(\rho,\phi,z)\), \(\Phi_{PW}(\rho,\phi,z)\) and \(\Phi_{N}(\rho,\phi,z)\) as needed. \\
The angular momentum (L) of the test particle about the central rotational axis of the halo remains conserved and can be used as one of the constraints in this problem. Due to the azimuthal symmetry inherent in this potential, the net effective potential experienced by a test particle within the asymmetric halo includes the contribution of the centrifugal force component  along the radial direction, given by $\biggl(\dfrac{L^2}{2{\rho}^2}\biggr)$. 
Thus the effective potential becomes: 
\begin{align} \label{effective_Pot}
  V =\, & \Phi_g+ \frac{L^2}{2{\rho}^2}\nonumber\\
     \implies V =\,& \Phi(\rho,\phi,z) + Dz + \frac{Q}{2}(2z^2-{\rho}^2) \nonumber \\
    & + \frac{O}{2}(2z^3-3z{\rho}^2) + \frac{L^2}{2{\rho}^2}
\end{align}
\subsection{Newtonian dynamics}

The Hamiltonian of this system can be written as, 
\begin{equation}\label{Hamiltonian}
    H = \frac{{p_{\rho}}^2+{p_z}^2}{2}+\frac{L^2}{2{\rho}^2}+\Phi_g
\end{equation}

So the equation of motion can be written as
\begin{subequations} 
\begin{align} 
    \dot{\rho}&=p_{\rho}, \\
    \dot{p_{\rho}}&=-\frac{\partial V}{\partial {\rho}}, \\
    \dot{z}&=p_z, \\
    \dot{p_{z}}&=-\frac{\partial V}{\partial {z}};
\end{align} 
\label{eqn:Newt_eom}
\end{subequations}

Here we have taken $GM_{BH}=1$, where $M_{BH}$ is the mass of the central compact object and the velocity of light in free space, $c=1$.

The Hamiltonian in Eq.~(\ref{Hamiltonian}) does not have explicit time dependence. Thus the energy of the system must be a constant, given by,
\begin{equation}
    E_{mec}= \frac{p_{\rho}^2+{p_z}^2}{2}+\frac{L^2}{2{\rho}^2}+\Phi_g
\end{equation}
So applying these energy and angular momentum conservation conditions for this system, we can find the region where the test particle is restricted to move. That region is defined as, 
\begin{equation} \label{eqn:HillEq_Newt}
    E^2-1- \frac{L^2}{2{\rho}^2} - \Phi_g  \geq  0
\end{equation}
where, $E$ is $\sqrt{1+2E_{mec}}$ \\
\subsection{Special relativistic dynamics}
Considering that the velocity components may not be negligible in comparison to the speed of light (i.e., equal to 1 after scaling), the inclusion of special relativistic corrections become important in the formulation of the equations of motion. Therefore, an alternative formalism should be developed by introducing the special relativistic corrections to the existing equations. Theoretically, this refined formalism is expected to yield more generalized results. In the context of relativistic motion, we can write the non-dimensional Lagrangian for a relativistic particle with unit mass as, 
\begin{equation}\label{Eq:Lagrangian}
    \mathcal{L}=-\frac{1}{\gamma}-\Phi_g
\end{equation}
Here, $\Phi_g$ denotes the potential defined in the preceding section within the cylindrical coordinate system given by Eq.~(\ref{Eq:potentialPhig}), with $\gamma$ being a parameter expressed in the same coordinate system as: 
\begin{equation}\label{Eq:gamma}
    \gamma=\dfrac{1}{\sqrt{1-(\dot{\rho}^2+\rho^2\dot{\phi}^2+\dot{z}^2)}}
\end{equation}
The Lagrangian presented above allows us to define the energy ($E$) and the angular momentum ($L$) as,
\begin{equation}
    E=\gamma+\Phi_g
\end{equation}
and
\begin{equation}\label{Eq:AngularMomentum}
    L=\dfrac{\partial \mathcal{L}}{\partial \dot{\phi}}=\gamma\rho^2\dot{\phi}
\end{equation}
By substituting these expressions for \(\gamma\) (Eq.(\ref{Eq:gamma})) and \(\dot{\phi}\) (Eq.~(\ref{Eq:AngularMomentum})) in terms of \(E\) and \(L\) into the Lagrangian Eq.~(\ref{Eq:Lagrangian}), we derive the Lagrange's equations governing the radial coordinate \(\rho\) and the axial coordinate \(z\). These equations, which are also given in several literatures (e.g., \cite{Gueron2001}, \cite{sankha_Nag2017}), can be expressed as follows.
\begin{subequations}
    \begin{align}
        (\Phi-E)\Ddot{\rho}&=\frac{\partial\Phi}{\partial\rho}(1-\dot{\rho}^2)-\frac{\partial\Phi}{\partial z}\dot{\rho}\dot{z}-\frac{L^2}{(E-\Phi)\rho^3},\\
         (\Phi-E)\Ddot{z}&=\frac{\partial\Phi}{\partial z}(1-\dot{z}^2)-\frac{\partial\Phi}{\partial\rho}\dot{\rho}\dot{z}
    \end{align}
\end{subequations}
Maintaining the non-relativistic definitions of \(p_\rho\) and \(p_z\) unchanged for the sake of comparison, the equations of motion take the form:
\begin{subequations} 
\begin{align} 
    \dot{\rho}&=p_{\rho}, \\
    \dot{p_{\rho}}&= \frac{1}{\Phi_{g}-E} \left[ \frac{\partial {\Phi_g}}{\partial{\rho}}(1-{p_{\rho}}^2)- \frac{\partial {\Phi_g}}{\partial{z}} p_{z}p_{\rho}- \frac{L^2}{(E-\Phi_{g}){\rho}^3} \right], \\
    \dot{z}&=p_z, \\
    \dot{p_z}&= \frac{1}{\Phi_{g}-E} \left[  \frac{\partial {\Phi_g}}{\partial{z}}(1-{p_z}^2) - \frac{\partial {\Phi_g}}{\partial{\rho}} p_{z}p_{\rho} \right];
\end{align}
\label{eqn:Relt_eqom}
\end{subequations}
And the motion of the test particle restricted to the region is given by,
\begin{equation} \label{eqn:HillEq_Rel}
    {p_{\rho}}^2+ {p_z}^2+ \frac{L^2}{(E-\Phi_{g})^2{\rho}^2}+\frac{1}{(E-{\Phi_g})^2}=1
\end{equation}

Here we have scaled the velocity and length by dividing the velocity of light in vacuum and ${GM_{BH}}/{c^2}$ respectively, where $M_{BH}$ is the mass of the central compact object. Also we consider the mass of the test particle and $GM_{BH}$ to be unity for the ease of numerical simulation.
It is crucial to note that, much like the normalization of velocities and distances, all expressions for physical quantities have been adjusted to ensure dimensionless representation. In particular, the temporal parameter \(t\) has been rescaled as \({ct}/{(r_g)}\), the momentum \(p\) has undergone scaling as \({p}/{(\mu c)}\) and the angular momentum \(L\) has been adjusted to \({L}/{(\mu cr_g)}\), where \(\mu\) signifies the mass of the test particle, which is assumed to be unity in this analysis.


\section{Poincar\'e Sections}

In this section, we employ the Poincaré Section (PS) technique to analyze the dynamical behaviour of our system. This method captures a lower-dimensional representation of the phase space by plotting intersections between the system's trajectory and a fixed plane. The test particle is confined to the $z=0$ plane, governed by constant angular momentum $(L)$ and energy $(E)$ within a 4-dimensional phase space.

The PS is constructed by identifying points of intersection between the test particle's trajectory and a 2-dimensional fixed plane along a specific direction on the 3-dimensional tour. Isolated points on the map correspond to periodic orbits, while chaotic orbits manifest as a scattered distribution of points. Additionally, stable quasi-periodic orbits (referred to as sticky orbits in our system) exhibit systematic patterns on the map, aligning with concentric Kolmogorov-Arnold-Moser (KAM) tori.

Qualitatively, the PS provides insight into system chaoticity, with increased chaos indicated by a reduction in regular pattern islands relative to scattered points. This approach facilitates the exploration of correlations between the degree of chaos and various dynamical parameters, where heightened chaos corresponds to diminishing regular orbit islands.

The Poincaré section has been employed as a diagnostic tool to elucidate the manifestation of chaos in celestial systems, such as the orbital motion of a particle within a halo, as documented in earlier scholarly works. Notably, some authors (e.g., \cite{vieira1996Henon_HeilDipHaloGR}, \cite{Vieira1999CoreShellGR}, \cite{Gueron2002}) have utilized this methodology within the framework of general relativity, while other (e.g., \cite{Ying2012}) have applied the technique exclusively within the Newtonian formalism. Additional studies conducted by  several authors (\cite{Gueron2001}, \cite{sankha_Nag2017} and \cite{saikat_das_SRC2023}) have extended the application of the Poincaré section to both Newtonian and Special relativistic dynamics. In the current investigation, we have utilized this approach to analyze both Newtonian and Special relativistic dynamics scenarios.

In our work all the Poincaré sections were generated on the $z = 0$ plane, with the orbits traverse away from the plane along their trajectories in the phase space ($p_z > 0$). Varied initial conditions, including coordinates ${\rho_0}(t = 0)$, $z_0(t = 0)$, and momentum $p_{\rho0}(t = 0)$, were explored for a specific set of orbital parameters. The initial $p_{z0}(t = 0)$ value was determined through conservation equation Eq.~(\ref{eqn:HillEq_Newt}) and Eq.~(\ref{eqn:HillEq_Rel}) for Newtonian and special relativistic corrected trajectories.

Numerical integration of the equations of motion, described by Eq.~(\ref{eqn:Newt_eom}) and Eq.~(\ref{eqn:Relt_eqom}), was conducted using MATLAB's ode45 module a widely utilized computational tool for solving ordinary differential equations. ode45 employs a variable-step, explicit Runge-Kutta method, adapting the time step for accurate numerical integration. This method's versatility makes it well-suited for precise simulations of complex astronomical dynamics. 

Consistency in our simulations has been maintained by selecting specific values for two fundamental constants, namely \(L = 4.2\) and \(E = 0.976\). We have investigated the Poincaré section for various sets of E and L values, observing rich and interesting chaotic dynamics for E values between 0.95 and 0.99 and L values between 3.5 and 6.7 in Newtonian dynamics. For the special relativistic case, the range of values of E and L for which interesting features were observed is smaller compared to the Newtonian case. Across these different sets of values, we have generally found that the poincare section shows a similar trend,  which is consistent with an earlier study by Viera et al. (1999) \cite{Vieira1999CoreShellGR}. Thus to study and understand the system's dynamical features in more detail, we have used the Smaller Alignment Index (SALI), which is computationally far more intensive and time-consuming for a representative value of E and L lying within the earlier stated range, which has been also widely used by earlier authors \cite{sankha_Nag2017}, \cite{Gueron2001} These constants have been applied uniformly across all simulations conducted in our study, ensuring a standardized and comparable basis for the analysis of system dynamics.


\begin{figure*}
        \begin{subfigure}{0.49\textwidth}
        \includegraphics[width=\textwidth]{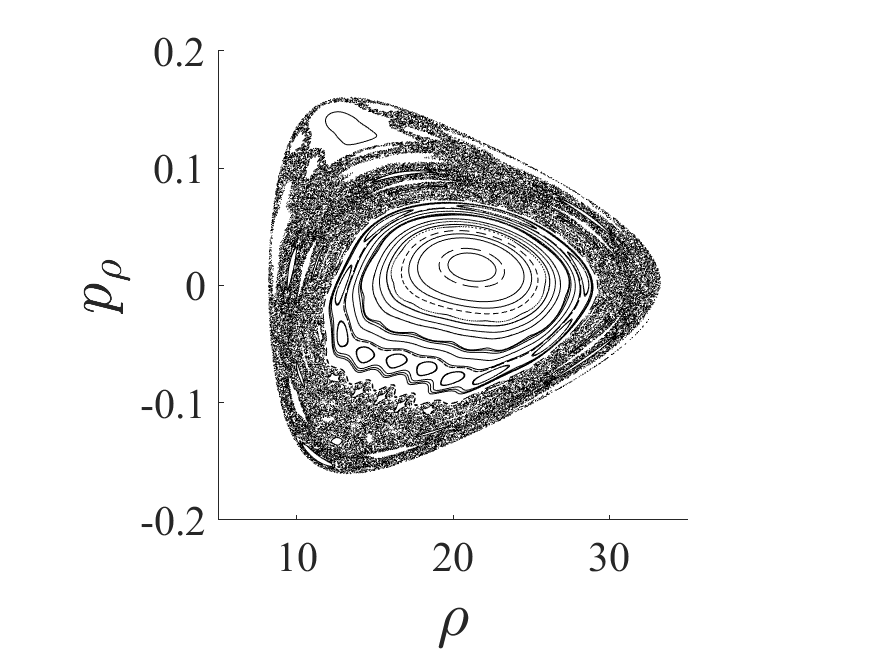}
        \caption{$a=0.2$, with out special relativistic correction}
        \label{fig:poincareN_D+O_spin0.2}
    \end{subfigure}
    \hfill
        \begin{subfigure}{0.49\textwidth}
        \includegraphics[width=0.7\textwidth]{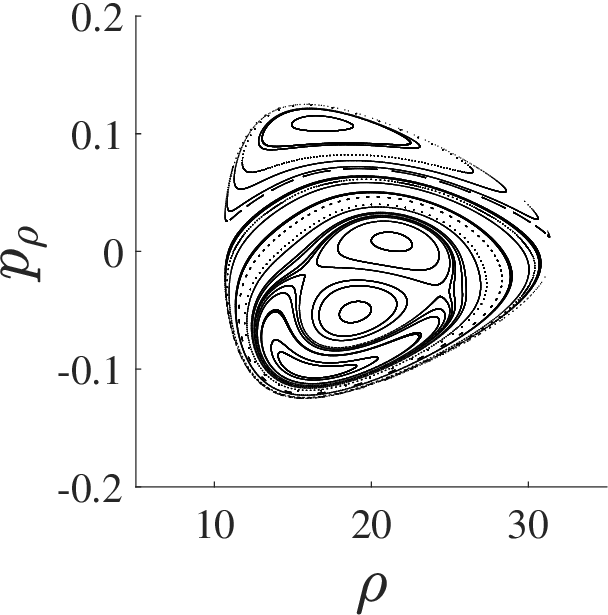}
        \caption{$a=0.8$, with out special relativistic correction}
        \label{fig:poincareN_D+O_spin0.8}
    \end{subfigure}
    \hfill
        \begin{subfigure}{0.49\textwidth}
        \includegraphics[width=\textwidth]{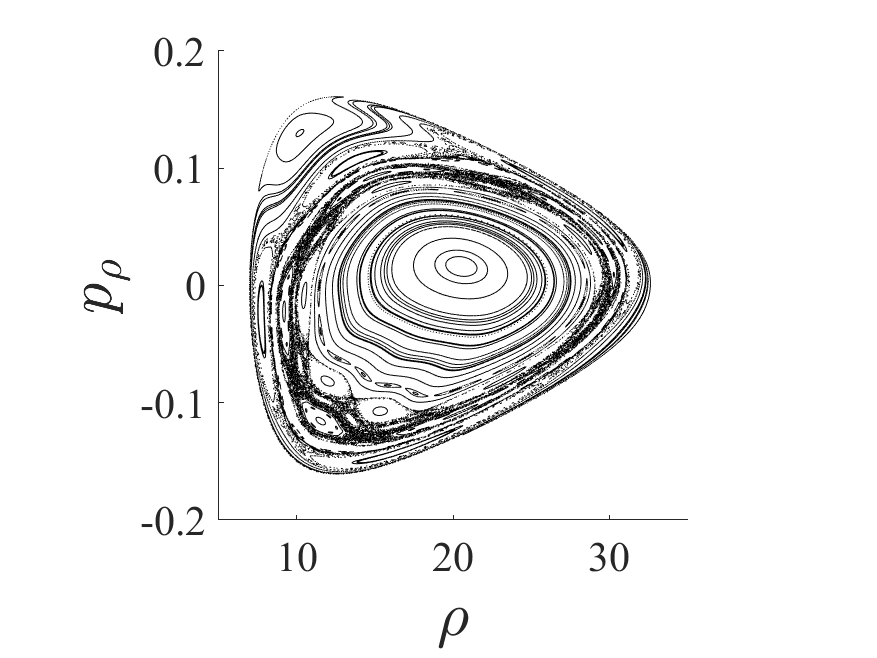}
        \caption{$a=0.2$, with special relativistic correction}
        \label{fig:poincareSR_D+O_spin0.2}
    \end{subfigure}
    \hfill
        \begin{subfigure}{0.49\textwidth}
        \includegraphics[width=0.7\textwidth]{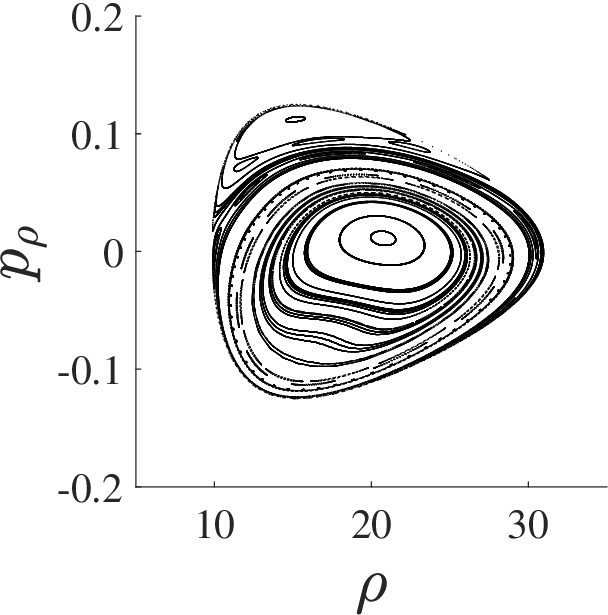}
        \caption{$a=0.8$, with special relativistic correction}
        \label{fig:poincareSR_D+O_spin0.8}
    \end{subfigure}
    \captionsetup{justification=raggedright,singlelinecheck=false}
    \caption{Poincar\'e maps on the cross-sectional plane $z = 0$ for the $D+O$ system with $D=2\times10^{-4}$, $O=2.4\times10^{-7}$ , $E = 0.976$, and $L = 4.20$. (a) and (b) without any special relativistic correction. (c) and (d) after special relativistic correction }
    \label{fig:poincare_D+O}
\end{figure*}

\begin{figure*}
        \begin{subfigure}[b]{0.49\linewidth}
        \centering
        \includegraphics[width=\textwidth]{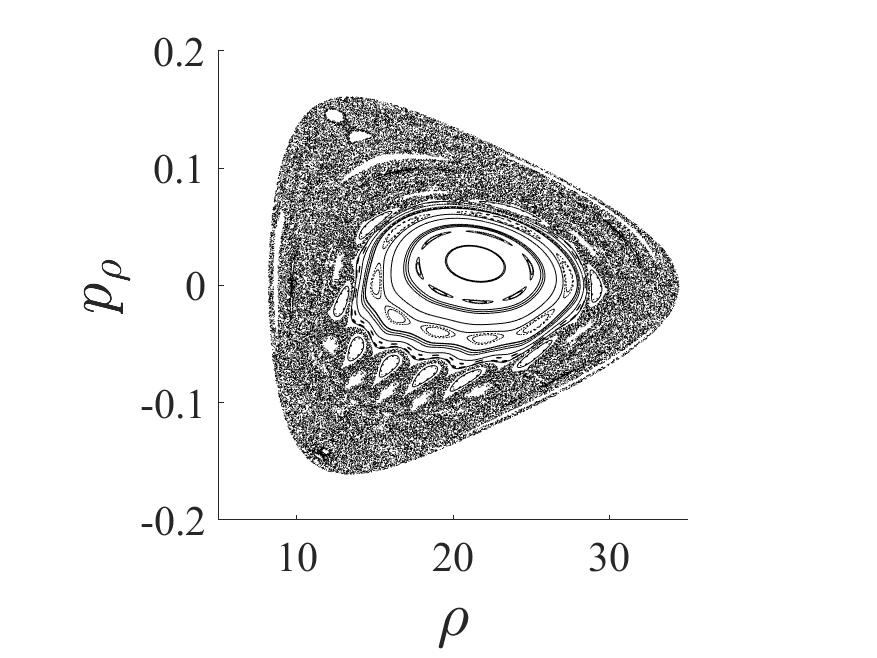}
        \caption{$a=0.2$, with out special relativistic correction}
        \label{fig:poincareN_D+Q+O_spin0.2}
    \end{subfigure}
    \hfill
        \begin{subfigure}[b]{0.49\linewidth}
        \centering
        \includegraphics[width=0.7\textwidth]{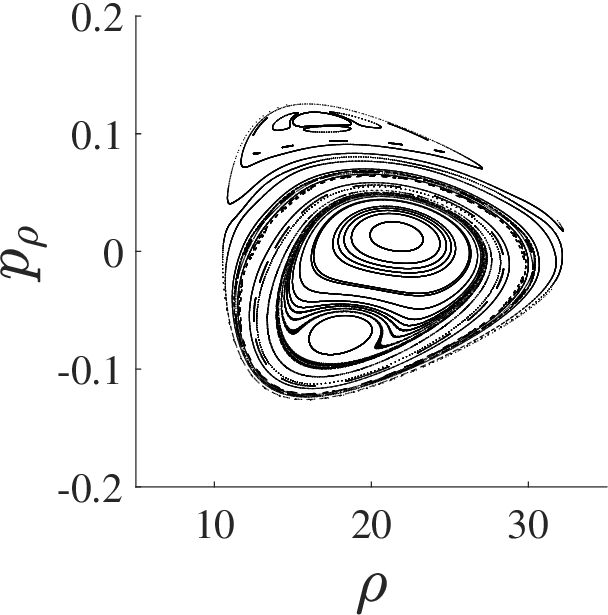}
        \caption{$a=0.8$, with out special relativistic correction}
        \label{fig:poincareN_D+Q+O_spin0.8}
    \end{subfigure}
    \hfill
        \begin{subfigure}[b]{0.49\linewidth}
        \centering
        \includegraphics[width=\textwidth]{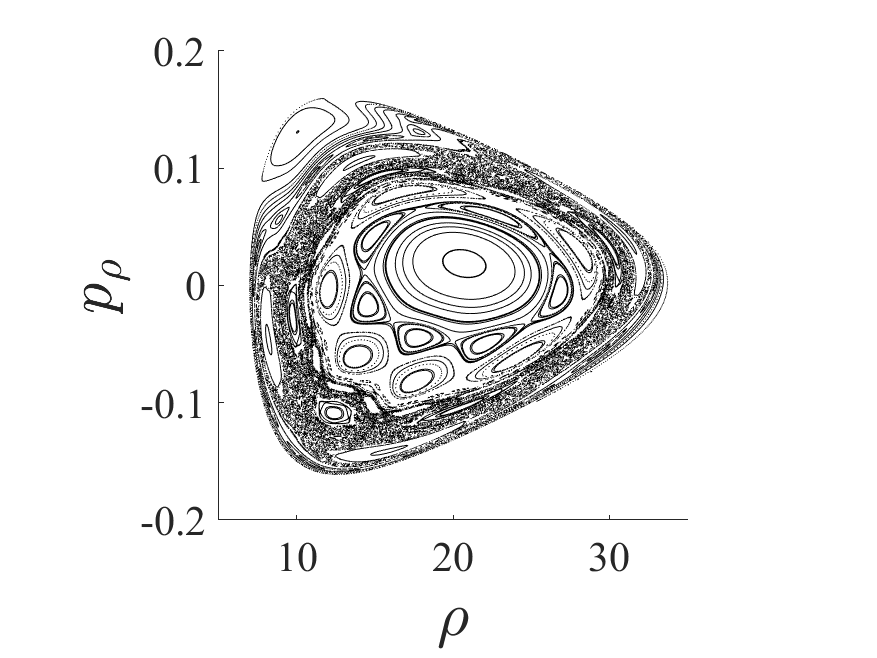}
        \caption{$a=0.2$, with special relativistic correction}
        \label{fig:poincareSR_D+Q+O_spin0.2}
    \end{subfigure}
    \hfill
        \begin{subfigure}[b]{0.49\linewidth}
        \centering
        \includegraphics[width=0.7\textwidth]{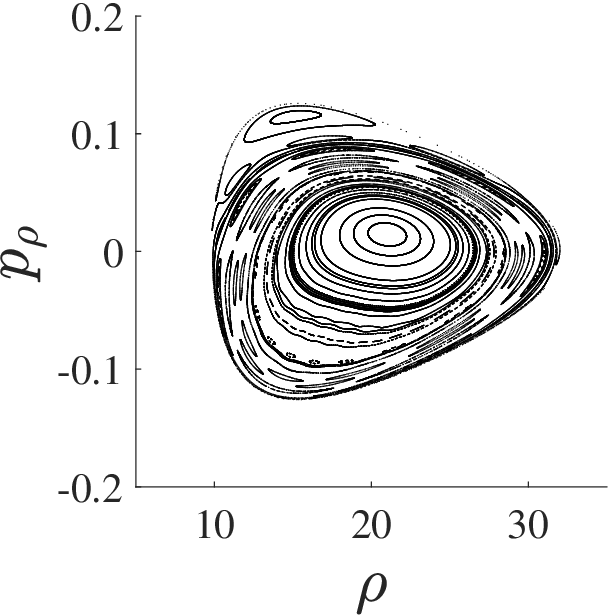}
        \caption{$a=0.8$, with special relativistic correction}
        \label{fig:poincareSR_D+Q+O_spin0.8}
    \end{subfigure}
    \captionsetup{justification=raggedright,singlelinecheck=false}
    \caption{Poincar\'e maps on the cross-sectional plane $z = 0$ for the $D+Q+O$ system with $D=2\times10^{-4}$, $Q=1\times10^{-6}$, $O=2.2\times10^{-7}$, $E = 0.976$, and $L = 4.20$. (a) and (b) without any special relativistic correction. (c) and (d) after special relativistic correction}
    \label{fig:poincare_D+Q+O}
\end{figure*}

\subsection{Newtonian dynamics}

In our study, we have explored the influence of higher-order multipolar terms, up to the 3rd order, in the halo mass distribution. For the central monopole term, we employed the ABN potential, wherein setting the spin parameter \(a=0\) yields the PW potential, and \(a = 1\) reverts to the Newtonian potential. Our investigation involves switching between various multipolar terms, successfully replicating results from prior works.
The investigation of the impact of higher order multipolar moments on the system dynamics has revealed noteworthy findings. Specifically, the inclusion of dipole ($D$), quadrupole ($Q$), and octupole ($O$) moments in various combinations has been examined, leading to insightful observations.
By selectively activating different terms, such as the dipole term (\(D\neq 0\), $Q$\;=\;0, $O$\;=\;0), we replicated plots from Nag et al. (2017) \cite{sankha_Nag2017} for varying spin values (\(a=0, a=0.3, a=0.8\)). Similarly, by turning on dipolar ($D$) term only ($D$\(\neq 0\), Q\;=\;0, O\;=\;0), we reproduced Gu\'eron \& Letelier (2001) \cite{Gueron2001} results for \(a=0\). Activating the quadrupole ($Q$) term alone ($D$\;=\;0, $Q$\(\neq 0\), $O$\;=\;0) replicated  findings of Ying \& Xin  (2012) \cite{Ying2012} for different \(a\). Setting \( a = 1 \) (Newtonian potential), various combination of multiolar terms upto the third order are used to replicate certain plots of Vieira \& Letelier (1999) for comparison \cite{Vieira1999CoreShellGR}.


 We systematically examined all possible combinations of these moments such as  \(D\neq0, Q=0,O=0\) ($D$-only), \(D=0, Q\neq0,O=0\) (Q-only), \(D=0, Q=0,O\neq0\) ($O$-only), \(D\neq0, Q\neq0,O=0\) ($D+Q$), \(D\neq0, Q=0,O\neq0\) ($D+O$), \(D=0, Q\neq0,O\neq0\) ($Q+O$), \(D\neq0, Q\neq0,O\neq0\) ($D+Q+O$) and observed that as the multipolar moments increase, chaos in the system intensifies. However, we noted a contrasting trend with increasing spin, where chaos decreases.
Thus our analysys indicates that systems featuring combinations such as $D+Q$, $D+O$, $Q+O$, and $D+Q+O$ exhibit heightened chaotic behaviour compared to systems characterized by individual multipolar moments with a rotating monopole compact object. Remarkably, the $D+Q+O$ configuration emerges as the most chaotic, followed by $D+O$, $D+Q$, and $Q+O$ arrangements. Notably, the introduction of the dipole moment appears to be a key contributor to the amplification of chaos within the system. Similar result was also found for the Newtonian potential by Vieira \& Letelier (1999) \cite{Vieira1999CoreShellGR}.
We have also observed that, with increasing spin, chaos decreases, regardless of the multipolar terms. This observation is consistent with findings in binary Kerr systems made by several other authers, De et al. (2020) \cite{SRC&SDey}. 

\subsection{Relativistic dynamics}
In this investigation, Poincaré sections were employed to analyze the dynamical evolution of systems using special relativistic formulation of the system, as opposed to Newtonian dynamics, as discussed in the preceding section. Our focus is centered on understanding the influence of multipolar terms and spin on the manifestation of chaos in these systems. Notably our examination consistent with findings by Gu\'eron \& Letelier (2001) \cite{Gueron2001} in the context of the PW potential with a dipole term, revealed that the special relativistic system exhibited greater chaos compared to its Newtonian counterpart.

Specifically, when exploring the impact of higher order multipolar terms on inducing chaos in the system, we observed some interesting trends. For individual multipolar terms and systems featuring combinations such as D+Q and Q+O, the dynamics of the special relativistic formulation of the system exhibited more chaos for PW potential.


In the case of ABN potential our study has indicated that the  Newtonian system exhibits motion patterns that closely  resemble the special relativistic system for slightly elevated value of angular momentum (L). This finding is in agreement with conclusion of Nag et al. (2017) \cite{sankha_Nag2017}. However when the angular momentum is reduced we see that the dynamics of this system tend towards the system with PW potential for individual multipolar terms and combinations such as $D+Q$, $Q+O$. 

In contrast, intriguing outcomes have emerged for configurations involving the dipolar ($D$) and Octupolar ($O$) terms such as D+O and $D+Q+O$. Notably, our observations indicate a distinctive pattern in the Poincaré sections, as shown in Fig.~(\ref{fig:poincare_D+O}) and Fig.~(\ref{fig:poincare_D+Q+O}), particularly for low spin values. Specifically, we observed an escalation of chaotic orbits in the Newtonian dynamics compared to its special relativistic counterpart for low spin value, which is contrary to observation of earlier author (\cite{Gueron2001}). However, with increasing spin values, a convergence in the orbital behaviours emerge indicating comparable motions in both frameworks, which is in accordance with earlier observation in the litterateur (\cite{sankha_Nag2017}).

\section{SALI}
In the last section, we have discussed a tool (PS) which only gives qualitative insight into the system's dynamics. However our subsequent aim is to discern the characteristics of our system  by distinguishing chaotic orbits and the degree of chaos from other kinds of orbits. The trajectories of test particles can be systematically categorized into two primary classes: bound and unbound or escaping orbits. To distinguish between these classes, a threshold parameter \(r_d\) has been introduced. If the trajectory satisfies the condition \(\sqrt{\rho^2+z^2} > r_d\), indicating orbits beyond this threshold, they are classified as unbound or escaping. Conversely, if the trajectory is below this threshold, the orbits are considered bound. For computational purposes, \(r_d\) has been set to 80.  The selection of this threshold has been made by observing orbits under various initial conditions. It has been noted that once the trajectory surpasses this chosen threshold value, no orbits remain bound. Within the bound orbits, further distinctions can be made among regular, sticky, and chaotic trajectories. \\
One of the efficient ways of classifying such orbits is by a tool called Smaller Alignment Index (SALI) (\cite{Skokos2001}, \cite{Skokos2004}). The SALI method relies on tracking the evolution of deviation vectors and assessing their alignment or misalignment with the system's most unstable direction, i.e. in the direction of the largest Lyapunov Characteristic Exponent (mLCE). In practice, we track an orbit’s time evolution alongside two normalized deviation vectors ($\Vec{x},\Vec{w_1},\Vec{w_2}$ with initial conditions $\vec{x}(0)$, $\vec{w_1}(0)$,$\vec{w_2}(0)$). Normalizing these vectors at each step prevents overflow issues due to exponential growth. 
We can determine these deviation vectors (\cite{Skokos2010}) by employing the following equations,
\begin{equation*}
    \dot{\vec{w}}=[J_{2N}.{{D^2}_{H}}(\vec{x}(t))].\vec{w} \text{, where}
\end{equation*}
\begin{equation*}
    J_{2N}= 
\begin{pmatrix}
    0_N & I_N \\
    -I_N & 0_N
\end{pmatrix}
\end{equation*}
 and 
\begin{equation*}
    {{D^2}_{H}(\vec{x}(t))}_{ij}=\frac{\partial^2H}{\partial{x}_i \partial{x}_j}{\Bigg|}_{\vec{x}(t)}, \qquad i,j=1,2,...,2N 
\end{equation*}
\\
SALI is defined as, 
\begin{align}
    \text{SALI}(t)=&\text{min}\Biggl\{ \Bigg\| \frac{\vec{w_1}(t)}{\big\|\vec{w_1}(t) \big\|} +\frac{\vec{w_2}(t)}{\big\|\vec{w_2}(t) \big\|} \Bigg\|, \nonumber \\
    &\Bigg\| \frac{\vec{w_1}(t)}{\big\|\vec{w_1}(t) \big\|} -\frac{\vec{w_2}(t)}{\big\|\vec{w_2}(t) \big\|} \Bigg\| \Biggr\}
\end{align}
where $\big\|.\big\|$ denotes the Euclidean norm of a vector.
Based on the above definition, we can conclude that for chaotic orbits, the deviation vectors tend to align with the mLCE and become either parallel or antiparallel. Consequently, for chaotic orbits, the SALI value equals zero. In contrast, for regular orbits, the deviation vectors become tangential to the torus on which the orbit is constrained, pointing in different directions. As a result, SALI fluctuates around a constant value for regular orbits. For computational purposes, threshold values have been selected to classify distinct subclasses of bound orbits as follows. Chaotic orbits are designated by SALI values exceeding $10^{-8}$, regular orbits are characterized by SALI values surpassing $10^{-4}$, and the intermediary category of sticky orbits falls within the range where the SALI value is bounded by values greater than $10^{-8}$ but less than $10^{-4}$.  \\

We have employed Newtonian dynamics to derive the deviation equations, which are subsequently used to calculate the Smaller Alignment Index (SALI) values. Specifically, the ABN potential was utilized exclusively for the determination of the SALI values.
\noindent
For our system, the deviation equations are expressed as follows,

\begin{subequations}
\begin{align}
    \dot{\delta{\rho}} & =\delta p_{\rho}, \\
    \dot{\delta{z}} &=\delta p_{z}, \\
    \dot{\delta{p_{\rho}}} & = \left(-\frac{\delta^2{\Phi}}{\delta{\rho}^2}+Q+3Oz-\frac{3L^2}{{\rho}^4}\right) {\delta \rho} + \nonumber \\ 
    & \left(-\frac{\delta^2{\Phi}}{\delta{\rho} \delta {z}}+3O\rho\right) \delta z, \\ 
    \dot{\delta{p_z}} & = \left(-\frac{\delta^2{\Phi}}{\delta {z} \delta{\rho}}+3O\rho\right) \delta \rho + \nonumber \\
    & \left(-\frac{\delta^2{\Phi}}{\delta {z}^2}-2Q-6Oz\right) \delta z
\end{align}
\label{eqn:variational_eqn}
\end{subequations}
\noindent
In every scenario, we have integrated the equations of motion for 90000 steps, using 1 time step =1 unit. This extended duration is deliberately chosen to ensure ample time for all orbits to unveil their inherent characteristics. Furthermore, for each simulation, a grid of initial conditions spanning 320 by 320 grid points is employed. This grid configuration serves as the spatial foundation for the systematic investigation of orbital behaviours within the chosen dynamical system. We have employed the ode45 tool within MATLAB for numerical integration of the equations of motion, given by Eq.~(\ref{eqn:Newt_eom}) and their corresponding variational equations Eq.~(\ref{eqn:variational_eqn}).
\\

\begin{figure*}
        \begin{subfigure}[b]{0.32\linewidth}
        \includegraphics[width=\textwidth]{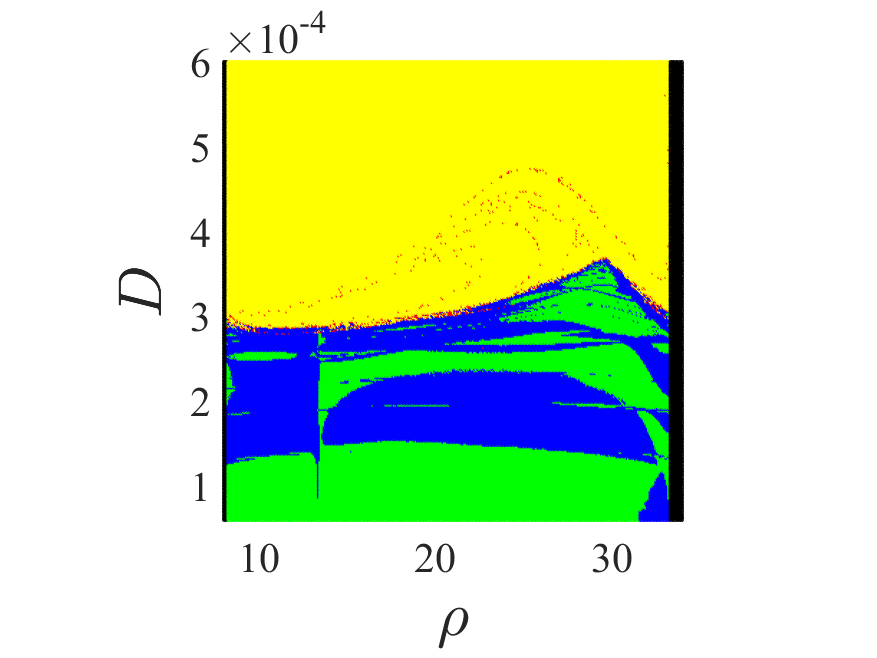}
        \caption{$a=0.2$}
        \label{fig:D-rpnane_spin02}
    \end{subfigure}
    \hfill
        \begin{subfigure}[b]{0.32\linewidth}
        \includegraphics[width=\textwidth]{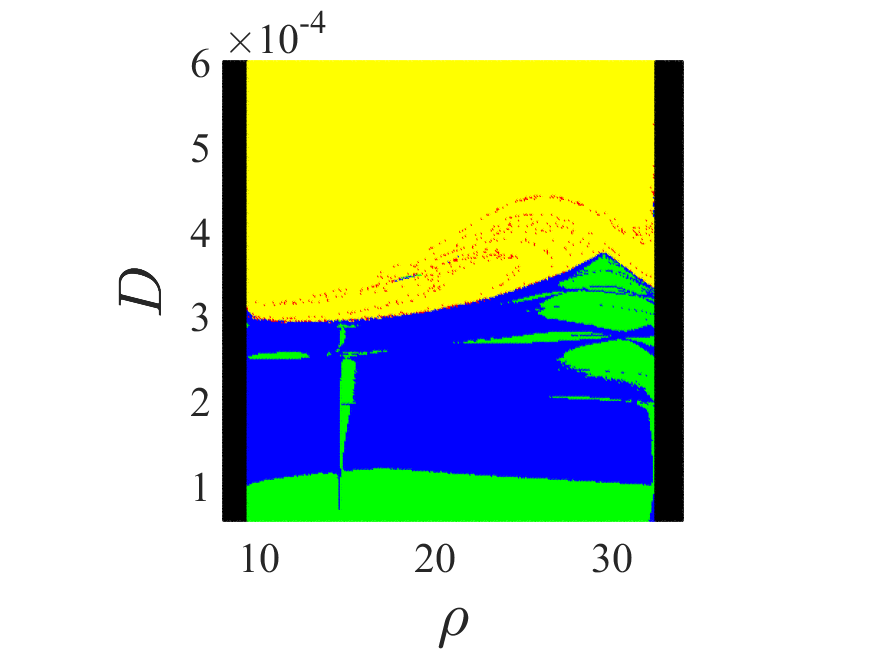}
        \caption{$a=0.5$}
        \label{fig:D-rpnane_spin05}
    \end{subfigure}
    \hfill
        \begin{subfigure}[b]{0.32\linewidth}
        \includegraphics[width=\textwidth]{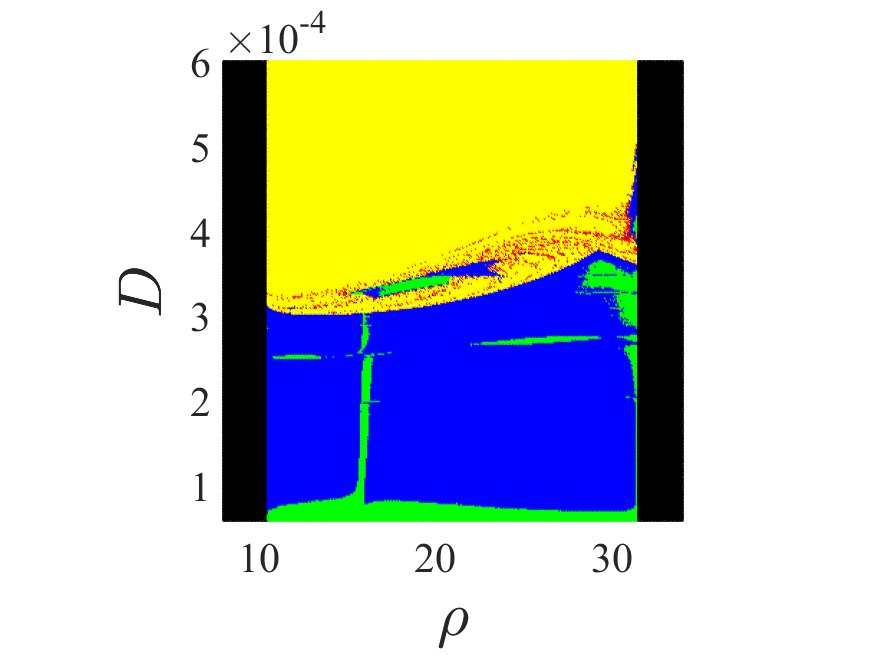}
        \caption{$a=0.8$}
        \label{fig:D-rpnane_spin08}
    \end{subfigure}
\captionsetup{justification=raggedright,singlelinecheck=false}
\caption{ Color-coded diagram on the $D-\rho$ plane with $Q=O=0$, E=0.976, $L=4.2$ for variation of spin. In this figure, the orbits are color-coded as follows: red indicates chaotic orbits, green represents regular orbits, blue denotes sticky orbits, and yellow signifies escaping orbits.}
    \label{fig:D-r_plane_SpinVariation}
\end{figure*}

\subsection{Orbit Classification}
We have used a scheme of colour coded diagram for classifying different kinds of orbits, wherein individual pixels are systematically assigned distinct colors corresponding to specific orbit types. The designated color scheme for various orbital categories is as follows: 
\\
1. Chaotic Orbits: Red. \\
2. Regular Orbits: Green. \\
3. Sticky Orbits: Blue. \\
4. Escaping or unbound Orbits: Yellow. \\
5. Disallowed Region: Black. \\
The use of this standardized color-coded approach has facilitated a rigorous and unambiguous representation of the orbital dynamics of our system throughout this work.

\subsubsection{The ($D-\rho$) plane} \label{SubSec:D-r}
In this subsection, we explore the influence of the dipolar moment ($D$) on orbits as a function of the scaled coordinate (\(\rho\)) and spin parameter ($a$). For this study, we have set  \( z = 0 \) and \( p_{\rho} = 0 \). Then we have calculated \( p_z \) using Eq.~(\ref{eqn:HillEq_Newt}) for every pair of \( \rho \) and \( D \) in a 320 $\times$ 320 grid ($\rho, D$) to generate all the plots in Fig.~(\ref{fig:D-r_plane_SpinVariation}). 

In Fig.~(\ref{fig:D-rpnane_spin02}) and Fig.~(\ref{fig:D-rpnane_spin08}) of the $D$-only system, the observed trend indicates a \(24\%\) increase in bound-to-unbound orbits as the spin value increases from $a=0.2$ to $a=0.8$. Consequently, sticky and chaotic orbits show a significant increase by \(99\%\) and \(153\%\), respectively. Conversely, there is a \(73\%\) reduction in regular orbits with increasing spin values from $a=0.2$ to $a=0.8$ for this $D$-only system. This suggests that an increase in spin enhances the likelihood of obtaining bound orbits. Among these bound orbits, sticky orbits are more predominant, comprising \(44\%\) of the total possible orbits, followed by regular \((7\%)\) and chaotic orbits \((1\%)\).

An intriguing observation from Fig.~(\ref{fig:D-rpnane_spin05}) is the presence of small isolated islands of bound orbits within the unbounded region, located at coordinates (18.4, $34.78\times 10^{-5}$) and (32.53, $42.0\times10^{-5}$) for spin \(a = 0.5\). Notably, these islands were not evident for \(a = 0.2\). With a further increase in spin to \(a = 0.8\), these two islands significantly expand, as depicted in Fig.~(\ref{fig:D-rpnane_spin08}).


\subsubsection{The ($Q-\rho$) plane} \label{SubSec:Q-r}

\begin{figure*}
    \centering
        \begin{subfigure}[b]{0.32\linewidth}
        \centering
        \includegraphics[width=\textwidth]{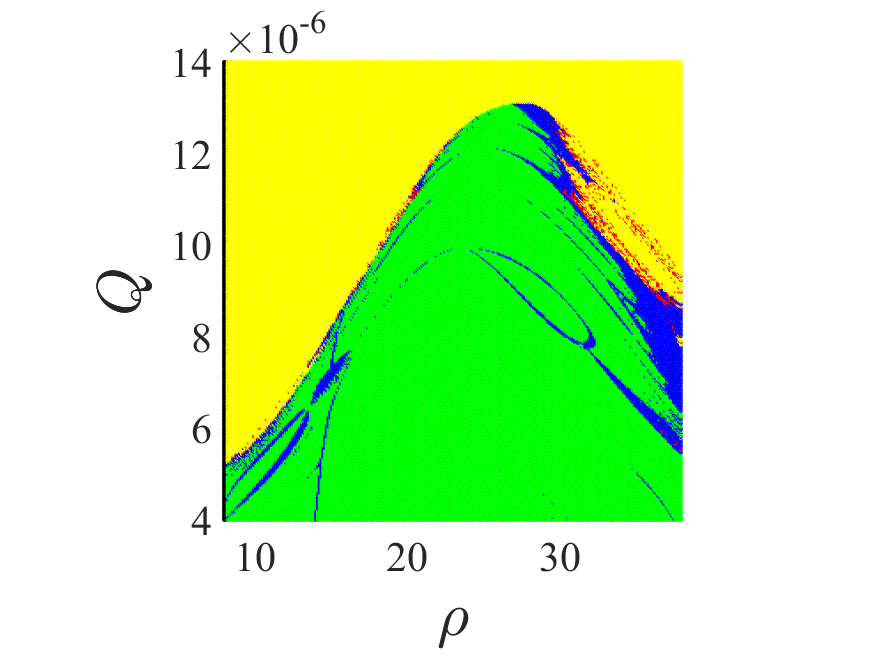}
        \caption{$D=0, a=0.2$}
        \label{fig:Q-r_Q_variation_spin02D0}
    \end{subfigure}
    \hfill
        \begin{subfigure}[b]{0.32\linewidth}
        \centering
        \includegraphics[width=\textwidth]{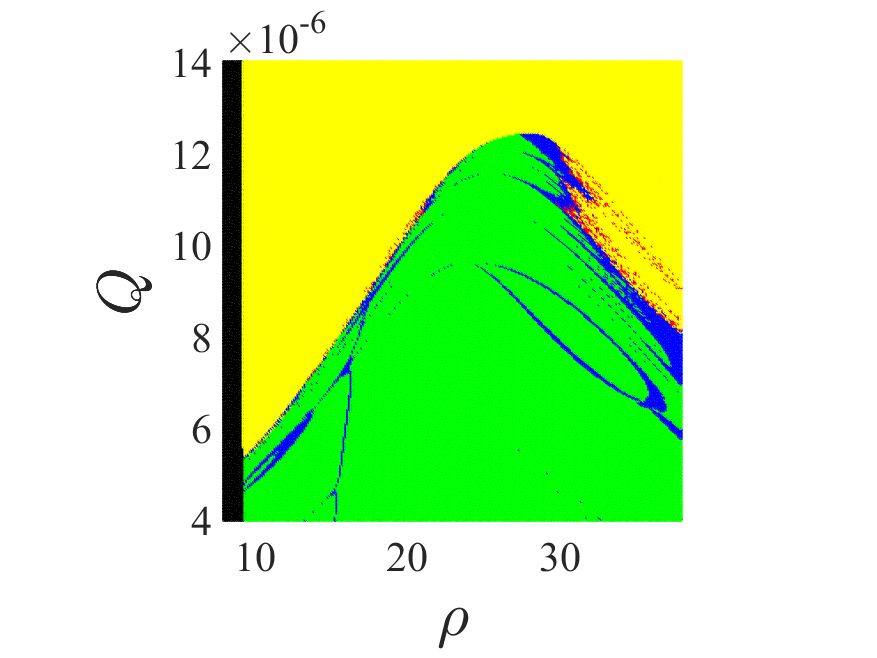}
        \caption{$D=0, a=0.5$}
        \label{fig:Q-r_Q_variation_spin05D0}
    \end{subfigure}
    \hfill
        \begin{subfigure}[b]{0.32\linewidth}
        \centering
        \includegraphics[width=\textwidth]{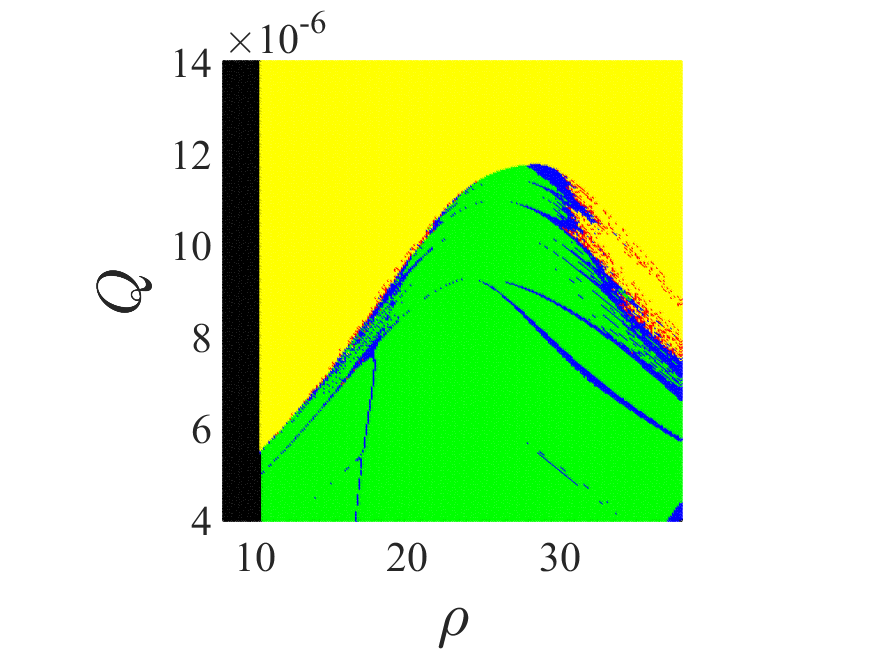}
        \caption{$D=0, a=0.8$}
        \label{fig:Q-r_Q_variation_spin08D0}
    \end{subfigure}
    \begin{subfigure}[b]{0.32\linewidth}
        \centering
        \includegraphics[width=\textwidth]{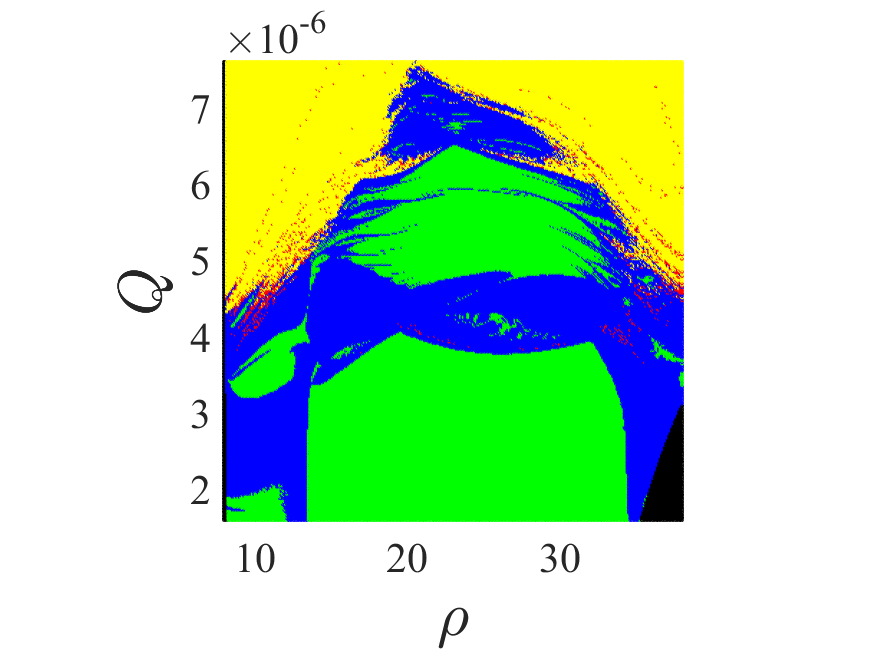}
        \caption{$D=2\times10^{-4}, a=0.2$}
        \label{fig:Q-r_Q_variation_spin02}
    \end{subfigure}
    \hfill
    \begin{subfigure}[b]{0.32\linewidth}
        \centering
        \includegraphics[width=\textwidth]{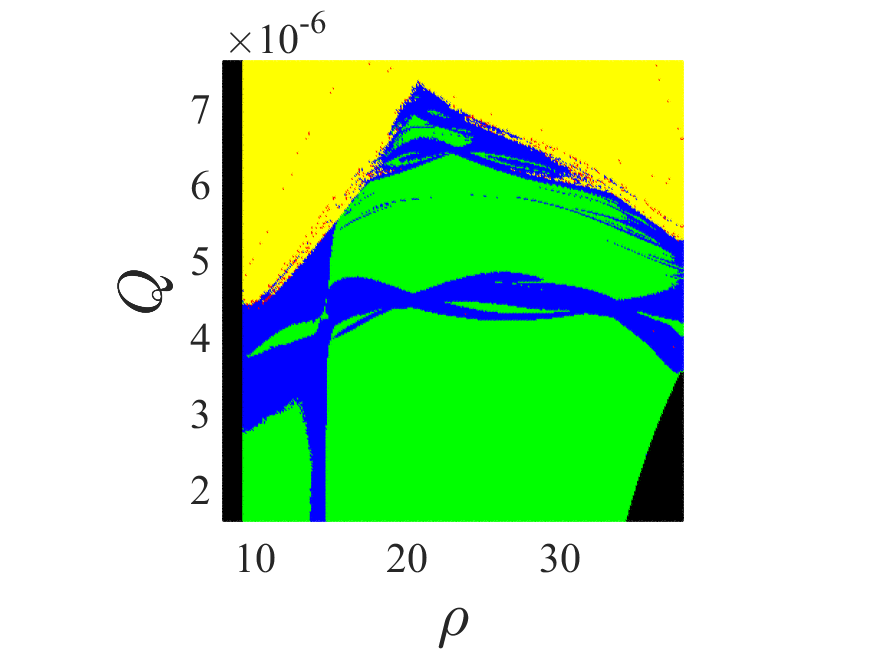}
        \caption{$D=2\times10^{-4}, a=0.5$}
        \label{fig:Q-r_Q_variation_spin05}
    \end{subfigure}
    \hfill
    \begin{subfigure}[b]{0.32\linewidth}
        \centering
        \includegraphics[width=\textwidth]{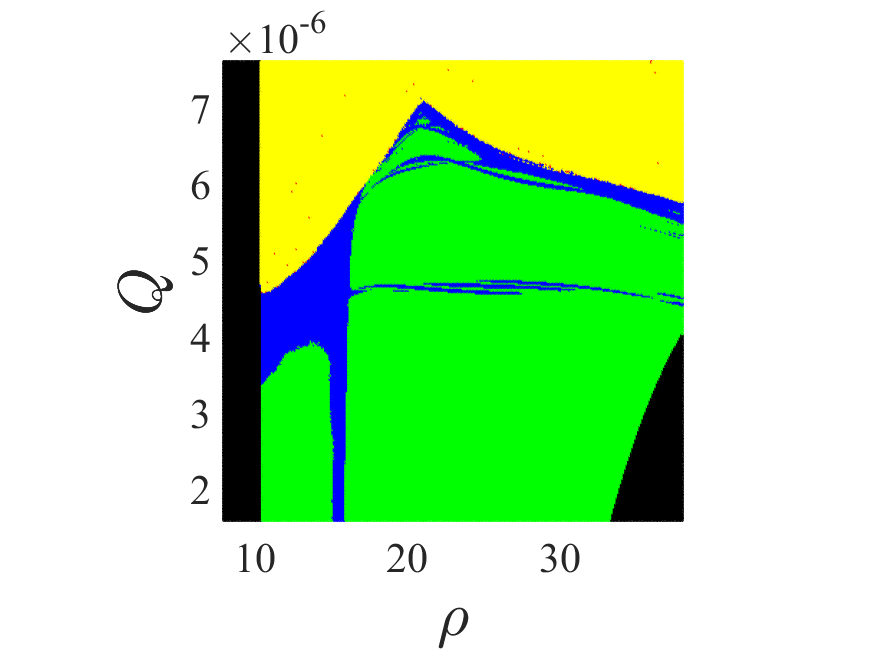}
        \caption{$D=2\times10^{-4}, a=0.8$}
        \label{fig:Q-r_Q_variation_spin08}
    \end{subfigure}
    \captionsetup{justification=raggedright,singlelinecheck=false}
    \caption{Color-coded diagram on the $Q-\rho$ plane with $O=0$, E=0.976, $L=4.2$ for variation of spin. In this figure, the orbits are color-coded as follows: red indicates chaotic orbits, green represents regular orbits, blue denotes sticky orbits, and yellow signifies escaping orbits.}
    \label{fig:Q-r_Q_variation}
\end{figure*}

Here we have investigated the dynamical influence of the quadrupolar moment ($Q$) as the function of the scaled coordinate ($\rho$) and the spin parameter ($a$) for the absence of  dipolar or octupolar term $(D\;=\;O\;=\;0)$ and when dipolar term is switched on 
($D=2\times10^{-4}$). In Figure~(\ref{fig:Q-r_Q_variation}), Fig.~(\ref{fig:Q-r_Q_variation_spin02D0}),  Fig.~(\ref{fig:Q-r_Q_variation_spin05D0}), and  Fig.~(\ref{fig:Q-r_Q_variation_spin08D0}) specifically elucidate the influence of the quadrupolar term and $\rho$ in the absence of other perturbing factors ($D=O=0, Q\neq0$, here after $Q$-only system) on the distribution of regular, sticky and chaotic orbits for different spin values. Similarly, sub-figures  Fig.~(\ref{fig:Q-r_Q_variation_spin02}),  Fig.~(\ref{fig:Q-r_Q_variation_spin05}) and Fig.~(\ref{fig:Q-r_Q_variation_spin08}) convey analogous information but specifically focus on the presence of the dipole term (\(O=0, D\neq0, Q\neq0\), here after $D+Q$ system). For this study, we set the values of \( z = 0 \) and \( p_{\rho} = 0 \). We then calculated the \( p_z \) value using Eq.~(\ref{eqn:HillEq_Newt}) for every pair of \( \rho \) and \( Q \) in a 320 $\times$ 320 grid ($\rho, Q$) to generate the plots.




In the absence of dipole or octupole terms ($D=O=0$), i.e. $Q$-only system, as illustrated in, Fig.~(\ref{fig:Q-r_Q_variation_spin02D0}), Fig.~(\ref{fig:Q-r_Q_variation_spin05D0}), and Fig.~(\ref{fig:Q-r_Q_variation_spin08D0}), our observations reveal that below a threshold value of $Q$ (denoted by $Q_t$), $Q_t<5.2 \times 10^{-6}$, $a=0.2$, all orbits remain bound across various scaled coordinate $\rho$. However, beyond a critical value $Q_c$ ($Q_c>13.1 \times 10^{-6}, a=0.2$), bound orbits cease to exist. This critical value is also spin dependent. As spin value increases, the critical value decreases i.e. $Q_c=11.8 \times 10^{-6}, a=0.8$. In the $D+Q$ system, when we switch on the dipole term in Fig.~(\ref{fig:Q-r_Q_variation_spin02}), Fig.~(\ref{fig:Q-r_Q_variation_spin05}) and Fig.~(\ref{fig:Q-r_Q_variation_spin08}), we observe a similar nature. However the critical and threshold values are smaller compared to the $Q$-only system. The sub-figures, Fig.~(\ref{fig:Q-r_Q_variation_spin02D0}), Fig.~(\ref{fig:Q-r_Q_variation_spin05D0}), and Fig.~(\ref{fig:Q-r_Q_variation_spin08D0}) also illustrate that with an increase in the scaled coordinate $\rho$, bound orbits systematically rise, peaking around $\rho\approx27$, and then progressively decline. When the dipole term is activated ($D\neq0,Q\neq0,O=0$), i.e. $D+Q$ system, a distinct declining pattern in bound orbits is observed compared to the $Q$-only system. Following the attainment of a maximum in the region of the bound orbits (denoted by green), they exhibit a gradual decrease, the descent being slower than in the case of $Q$ only system. Furthermore, we have observed that this slope is influenced by the value of D, with higher values of D leading to even smaller slopes.
In Fig.~(\ref{fig:Q-r_Q_variation_spin02D0}) and Fig.~(\ref{fig:Q-r_Q_variation_spin08D0}) of the $Q$-only system, the observed trend reveals a decrease in the bound-to-unbound orbit by a $20\%$ with increasing spin values from $a=0.2$ to $a=0.8$.  Interestingly, the presence of sticky and chaotic orbits is minimally influenced by variations in spin. Consequently, we observe \(10\%\) reduction in regular orbits with increasing spin values from $a=0.2$ to $a=0.8$ for this $Q$-only system roughly. 

Our findings suggest that incorporating the dipole term along with the quadrupole term significantly alters the inherent nature of this multipolar system with central compact rotating object. This observation can be realised from the detailed explanation of our analysis of Fig.~(\ref{fig:Q-r_Q_variation}), as described below.

We observe that in the $D+Q$ system, there is a considerable increase in sticky and chaotic orbits by  $304\%$ and $16\%$ respectively on activating the dipole term and for a low spin value of $a=0.2$, as compared to the $Q$-only system in Fig.~(\ref{fig:Q-r_Q_variation_spin02D0}) and Fig.~(\ref{fig:Q-r_Q_variation_spin02}). However, with an increase in spin to $a=0.8$ in Fig.~(\ref{fig:Q-r_Q_variation_spin08}), there is a notable reduction in sticky and chaotic orbits by \(68\%\) and  \(96\%\), respectively. Consequently, regular orbits show a significant increase of \(56\%\) for this $D+Q$ system, exhibiting an opposite trend compared to the $Q$-only system, where we have observed a decrease in the number of regular orbits with increasing spin.

\subsubsection{The ($O-\rho$) plane} \label{SubSec:O-r}
\begin{figure*}
    \centering
        \begin{subfigure}[b]{0.32\linewidth}
        \centering
        \includegraphics[width=\textwidth]{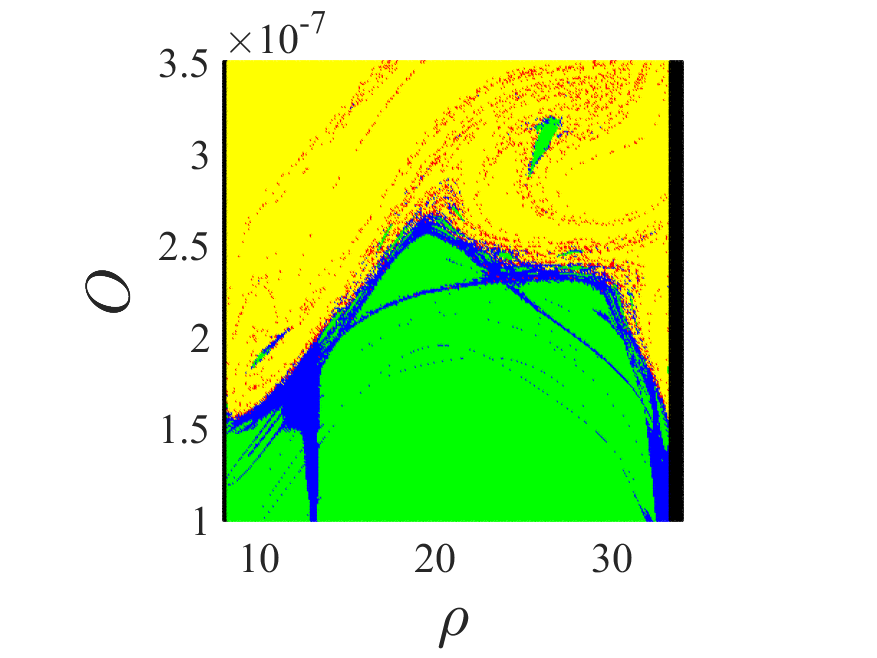}
        \caption{$D=0, a=0.2$}
        \label{fig:O-r_O_variation_spin02D0}
    \end{subfigure}
    \hfill
        \begin{subfigure}[b]{0.32\linewidth}
        \centering
        \includegraphics[width=\textwidth]{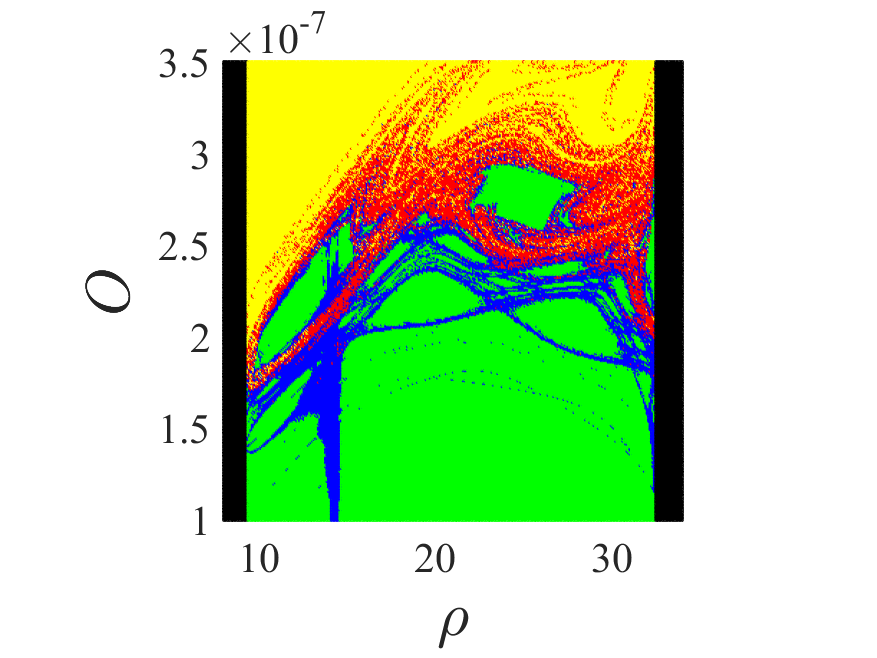}
        \caption{$D=0, a=0.5$}
        \label{fig:O-r_O_variation_spin05D0}
    \end{subfigure}
    \hfill
        \begin{subfigure}[b]{0.32\linewidth}
        \centering
        \includegraphics[width=\textwidth]{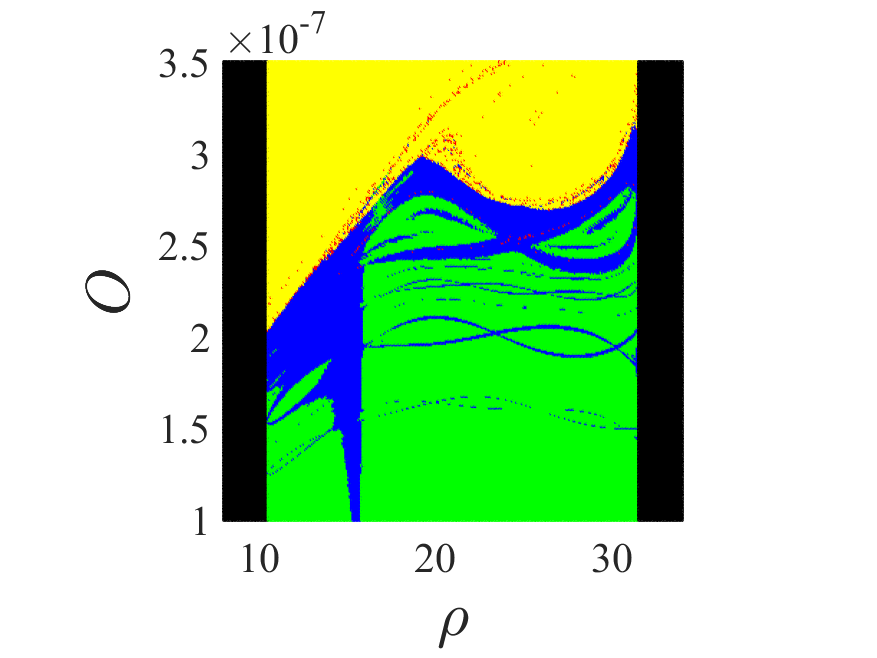}
        \caption{$D=0, a=0.8$}
        \label{fig:O-r_O_variation_spin08D0}
    \end{subfigure}
    \begin{subfigure}[b]{0.32\linewidth}
        \centering
        \includegraphics[width=\textwidth]{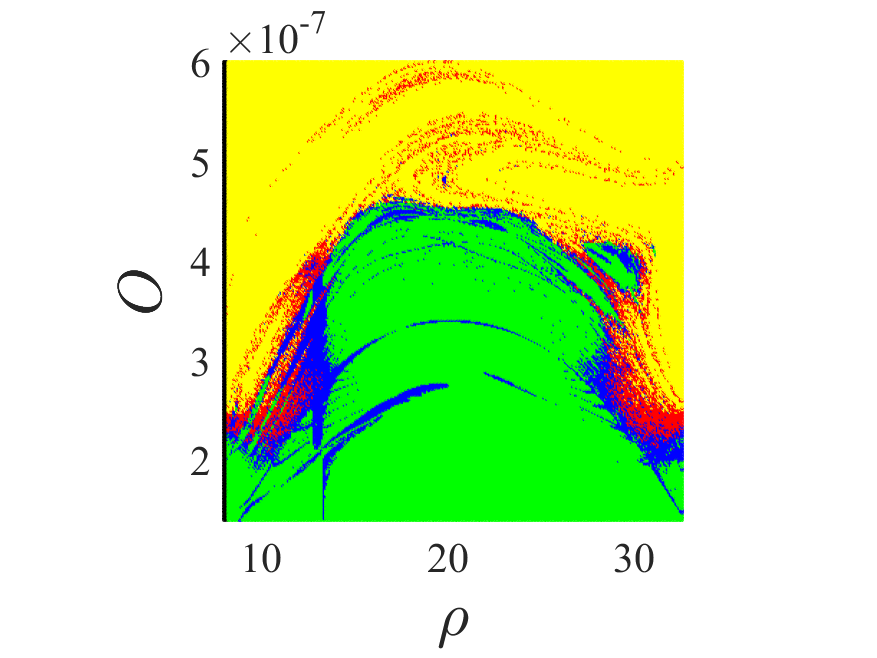}
        \caption{$D=2\times10^{-4}, a=0.2$}
        \label{fig:O-r_O_variation_spin02}
    \end{subfigure}
    \hfill
    \begin{subfigure}[b]{0.32\linewidth}
        \centering
        \includegraphics[width=\textwidth]{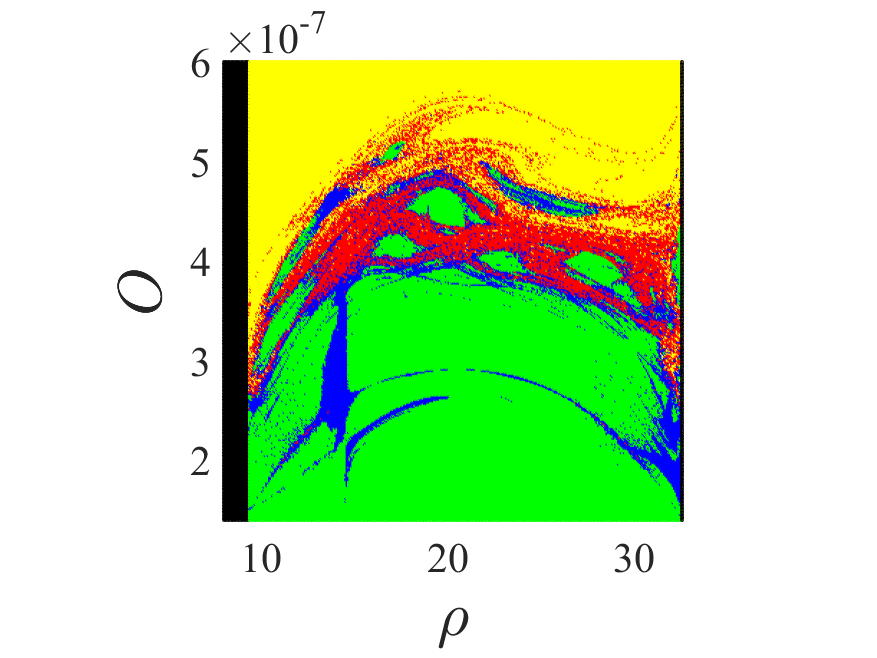}
        \caption{$D=2\times10^{-4}, a=0.5$}
        \label{fig:O-r_O_variation_spin05}
    \end{subfigure}
    \hfill
    \begin{subfigure}[b]{0.32\linewidth}
        \centering
        \includegraphics[width=\textwidth]{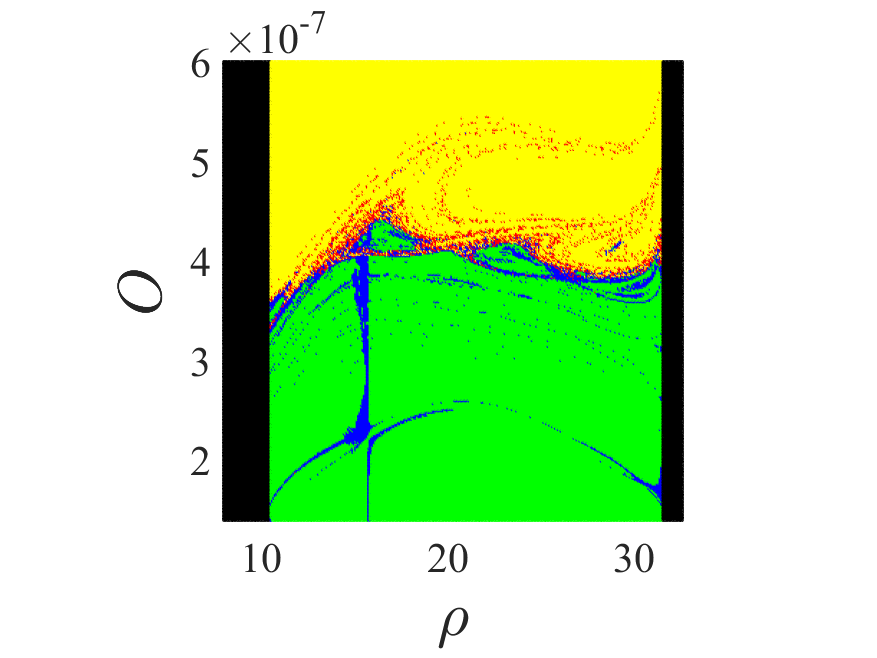}
        \caption{$D=2\times10^{-4}, a=0.8$}
        \label{fig:O-r_O_variation_spin08}
    \end{subfigure}
    \captionsetup{justification=raggedright,singlelinecheck=false}
    \caption{Color-coded diagram on the $O-\rho$ plane with $Q=0$, E=0.976, $L=4.2$ for the variation of spin. In the figure, the orbits are color-coded as follows: red indicates chaotic orbits, green represents regular orbits, blue denotes sticky orbits, and yellow signifies escaping orbits.}
    \label{fig:O-r_O_variation}
\end{figure*}

In this subsection, we investigated the dynamical impact of the octupole moment ($O$) as a function of scaled coordinate $\rho$ and spin parameter (a) under two scenarios: (i), when only octupole term is present (\(D=0\), \(Q=0\), \(O\neq0\), henceforth denoted by $O$-only  system) in Fig.~(\ref{fig:O-r_O_variation_spin02D0}), Fig.~(\ref{fig:O-r_O_variation_spin05D0}), and Fig.~(\ref{fig:O-r_O_variation_spin08D0}) and (ii) when the dipole and octupole terms are both present (\(D\neq0\), \(Q=0\), \(O\neq0\), henceforth denoted by D+O system) in Fig.~(\ref{fig:O-r_O_variation_spin02}), Fig.~(\ref{fig:O-r_O_variation_spin05}), Fig.~(\ref{fig:O-r_O_variation_spin08}). {For this study, we set the values of \( z = 0 \) and \( p_{\rho} = 0 \). We then calculated the \( p_z \) value using Eq.~(\ref{eqn:HillEq_Newt}) for every pair of \( \rho \) and \( O \) in a 320 $\times$ 320 grid ($\rho, O$) to generate the plots.

Taking into account all sub-figures in Fig.~(\ref{fig:O-r_O_variation}), we observed that the number of chaotic orbits initially increase with spin and reach a maximum value, after which they begin to decrease. Conversely, escaping orbits exhibit an opposite trend i.e. they decrease with spin and reach a minimum value, after which they begin to increase. Specifically, our analysis indicates that increasing spin from $0.2$ to $0.5$ results in an elevated chaotic-to-escaping orbit ratio (C/E), escalating by \(714\%\) for the $O$-only system and by \(180\%\) for the $D+O$ system. Further increasing spin to $a=0.8$ leads to a reduction in the  chaotic-to-escaping orbit ratio by \(95\%\) and \(83\%\) for the $O$-only and $D+O$ systems respectively.

Additionally in Fig.~(\ref{fig:O-r_O_variation}), it is observed that some regular and sticky isolated regions appear between the chaotic and escaping zones, particularly for low spin values. Notably, certain chaotic and escaping orbits surrounding these isolated regions transition into either regular or sticky orbits. Conversely, there is a steady increase in the count of regular and sticky orbits with increasing spin. 

Our findings indicate a similar behaviour for both $D+O$ and $O$-systems. The $D+O$ system exhibits  slightly more chaos compared to the $O$-only system at both low and high spin values. This implies that the inclusion of the dipole term with the octupole term does not significantly alter the inherent nature of the system, except for introducing more chaos.

\subsubsection{The ($p_{\rho}-\rho$) plane} \label{SubSec:p-r}


\begin{figure*}
        \begin{subfigure}[b]{0.32\linewidth}
        \centering
        \includegraphics[width=\textwidth]{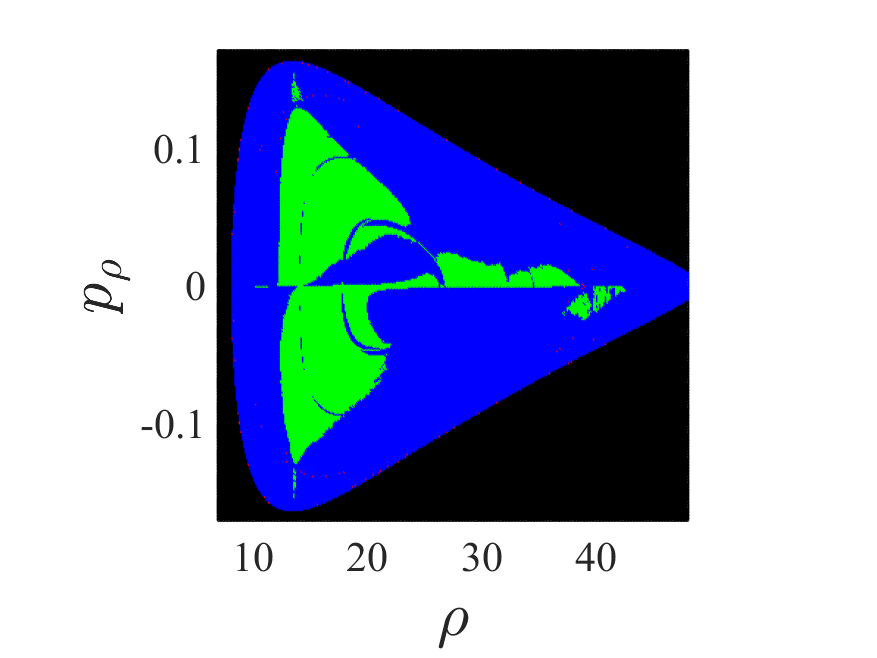}
        \caption{$a=0.2$, $Q=5.2\times10^{-6}$}
        \label{fig:p-r_plane_spin02D0O0}
    \end{subfigure}
    \hfill
        \begin{subfigure}[b]{0.32\linewidth}
        \centering
        \includegraphics[width=\textwidth]{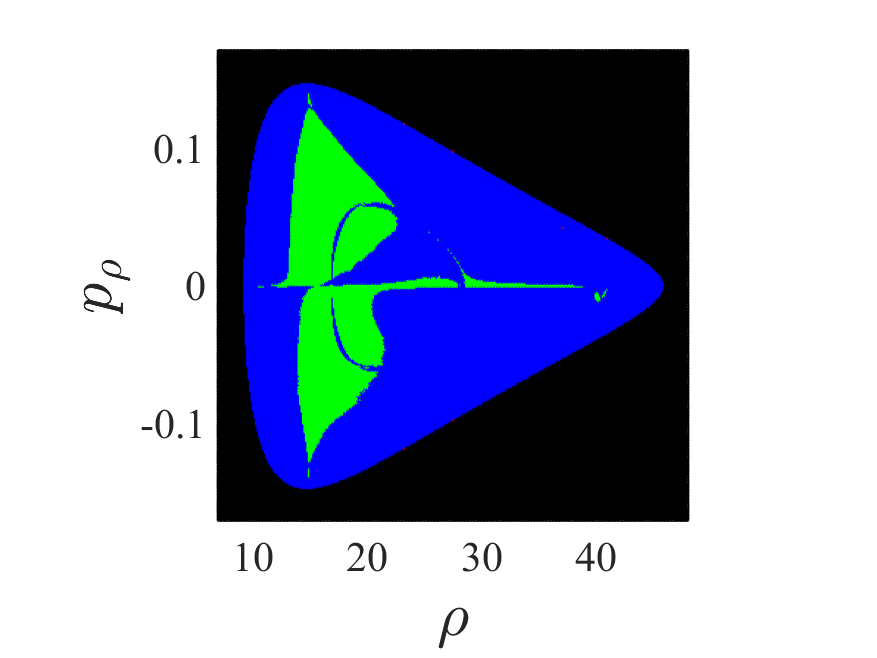}
        \caption{$a=0.5$, $Q=5.2\times10^{-6}$}
        \label{fig:p-r_plane_spin05D0O0}
    \end{subfigure}
    \hfill
        \begin{subfigure}[b]{0.32\linewidth}
        \centering
        \includegraphics[width=\textwidth]{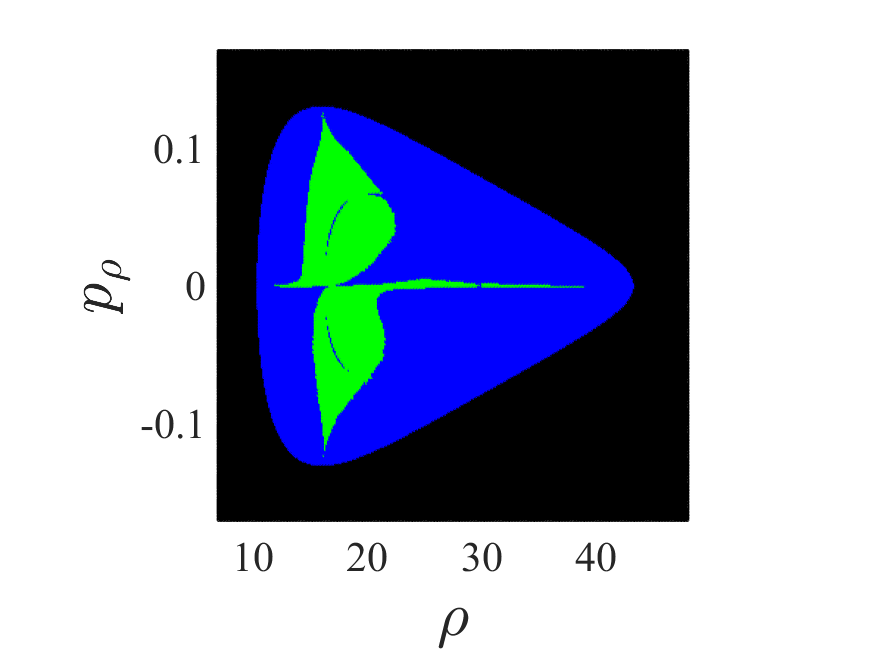}
        \caption{$a=0.8$, $Q=5.2\times10^{-6}$}
        \label{fig:p-r_plane_spin08D0O0}
    \end{subfigure}
    \hfill
        \begin{subfigure}[b]{0.32\linewidth}
        \centering
        \includegraphics[width=\textwidth]{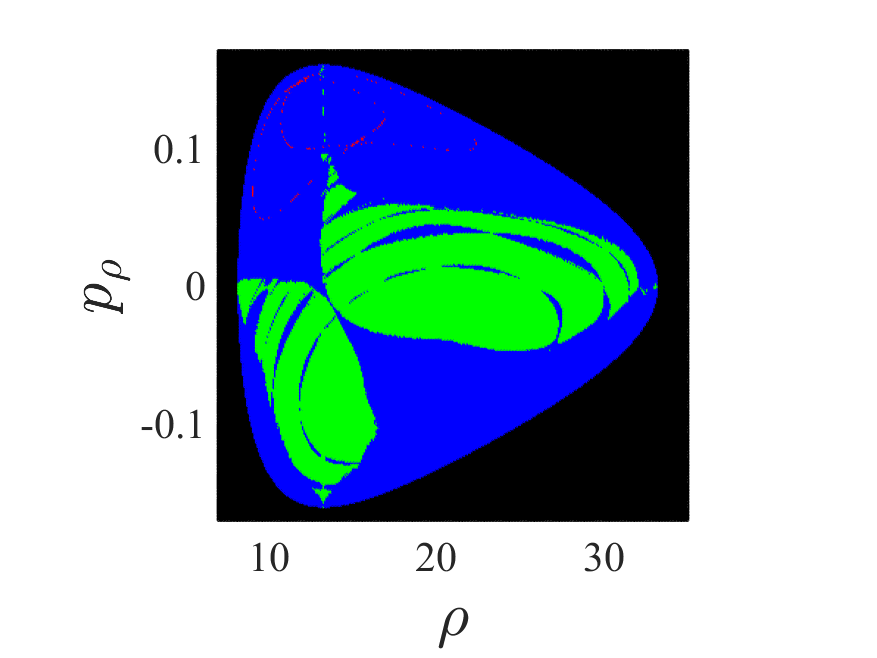}
        \caption{$a=0.2$, $O=1.4\times10^{-7}$}
        \label{fig:p-r_plane_spin02D0Q0}
    \end{subfigure}
    \hfill
        \begin{subfigure}[b]{0.32\linewidth}
        \centering
        \includegraphics[width=\textwidth]{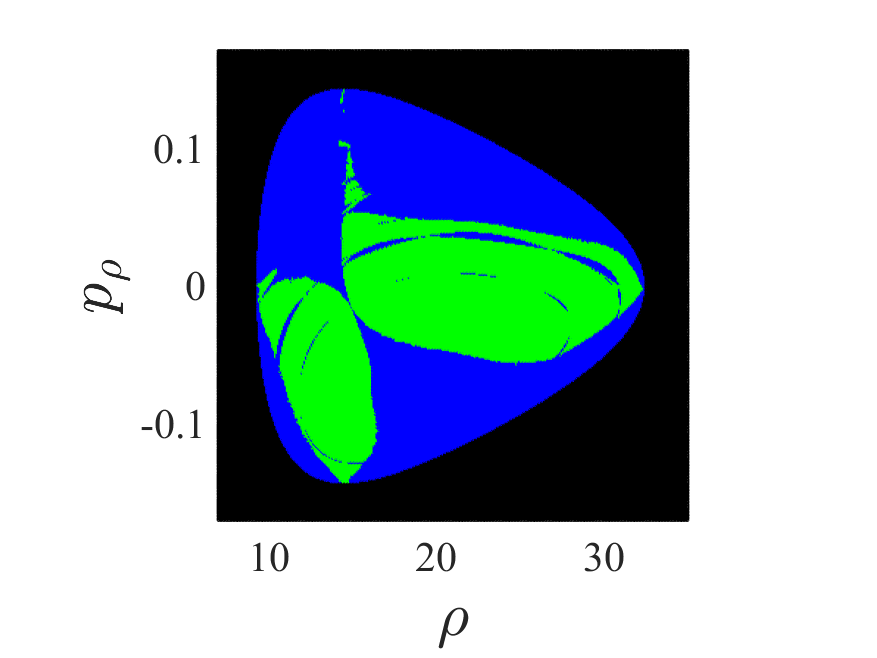}
        \caption{$a=0.5$, $O=1.4\times10^{-7}$}
        \label{fig:p-r_plane_spin05D0Q0}
    \end{subfigure}
    \hfill
        \begin{subfigure}[b]{0.32\linewidth}
        \centering
        \includegraphics[width=\textwidth]{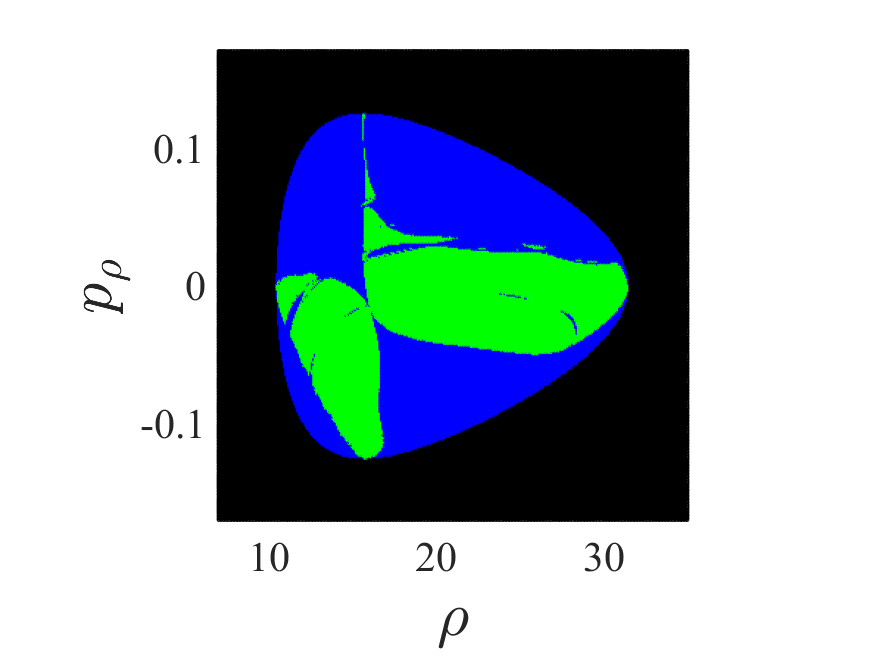}
        \caption{$a=0.8$, $O=1.4\times10^{-7}$}
        \label{fig:p-r_plane_spin08D0Q0}
    \end{subfigure}
    \hfill
        \begin{subfigure}[b]{0.32\linewidth}
        \centering
        \includegraphics[width=\textwidth]{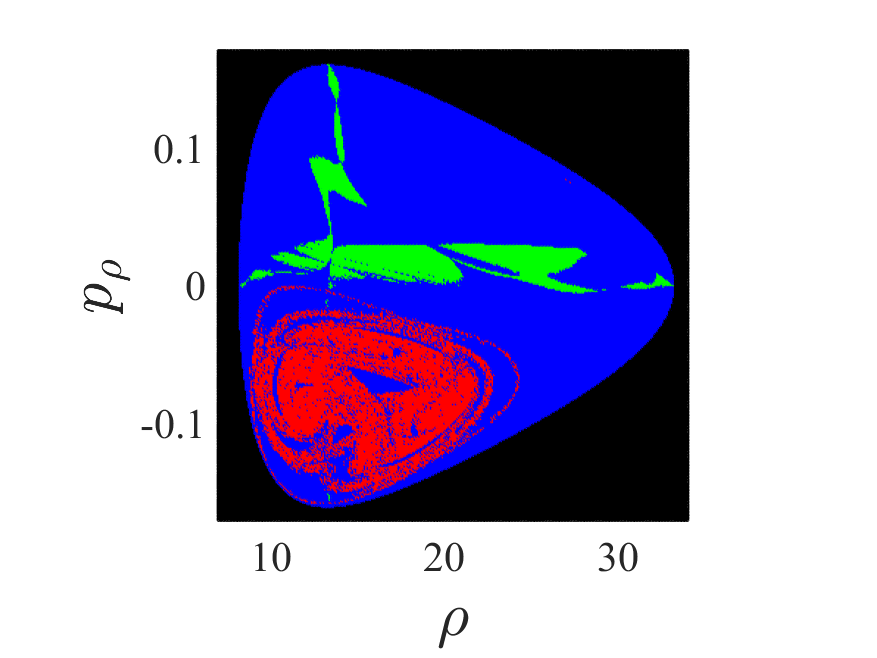}
        \caption{$a=0.2$, $D=2.8\times10^{-4}$}
        \label{fig:p-r_plane_spin02OnlyD}
    \end{subfigure}
    \hfill
        \begin{subfigure}[b]{0.32\linewidth}
        \centering
        \includegraphics[width=\textwidth]{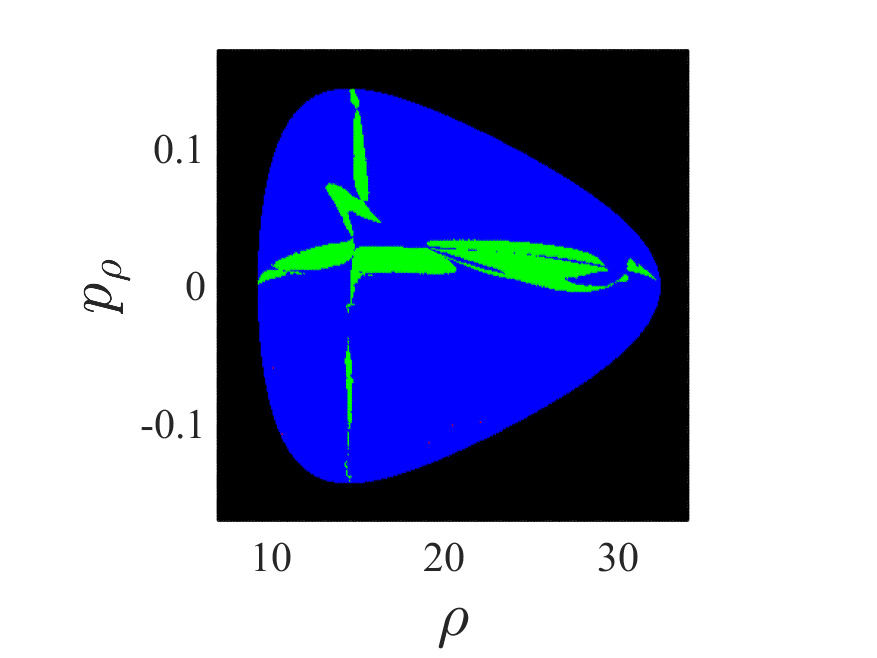}
        \caption{$a=0.5$, $D=2.8\times10^{-4}$}
        \label{fig:p-r_plane_spin05OnlyD}
    \end{subfigure}
    \hfill
        \begin{subfigure}[b]{0.32\linewidth}
        \centering
        \includegraphics[width=\textwidth]{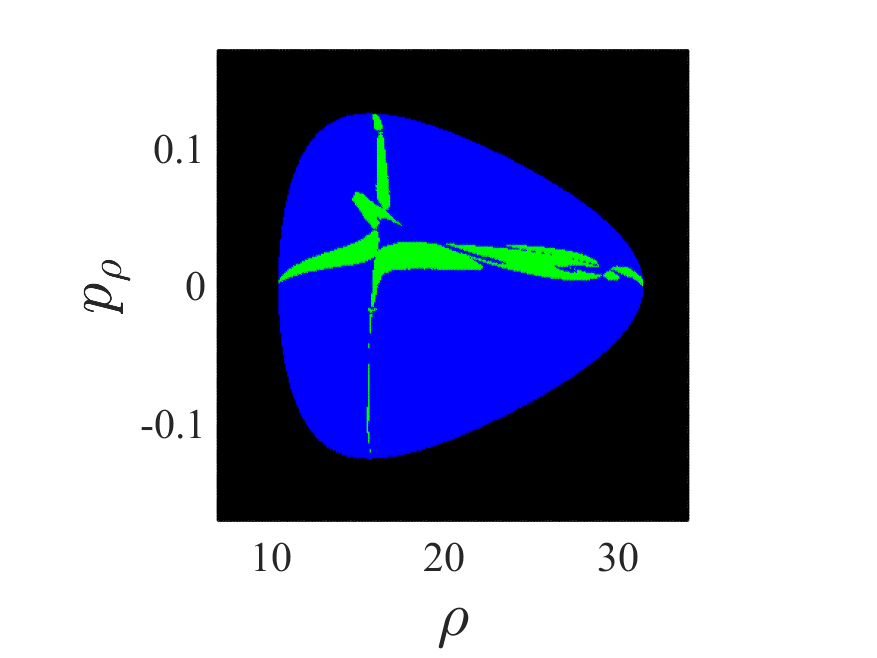}
        \caption{$a=0.8$, $D=2.8\times10^{-4}$}
        \label{fig:p-r_plane_spin08OnlyD}
    \end{subfigure}
        \hfill
    \captionsetup{justification=raggedright,singlelinecheck=false}
    \caption{ Color-coded diagrams in the $p_{\rho}-\rho$ plane depict systems with varying spin parameters: $Q$-only system ((\ref{fig:p-r_plane_spin02D0O0}), (\ref{fig:p-r_plane_spin05D0O0}), (\ref{fig:p-r_plane_spin08D0O0})); $O$-only system ( (\ref{fig:p-r_plane_spin02D0Q0}), (\ref{fig:p-r_plane_spin02D0Q0}), (\ref{fig:p-r_plane_spin02D0Q0})); $D$-only system ( (\ref{fig:p-r_plane_spin02OnlyD}), (\ref{fig:p-r_plane_spin05OnlyD}), (\ref{fig:p-r_plane_spin08OnlyD})), respectively. $L=4.2$ and $E=0.976$ are same for all the plots. In the figure, the orbits are color-coded as follows: red indicates chaotic orbits, green represents regular orbits and blue denotes sticky orbits.}
    \label{fig:p-r_plane_OnlyMP_spin_variation}
\end{figure*}

 Color-coded diagrams in the $p_{\rho}-\rho$ plane depict systems with varying spin parameters:

\begin{figure*}
\centering
        \begin{subfigure}[b]{0.32\linewidth}
        \centering
        \includegraphics[width=2.2in]{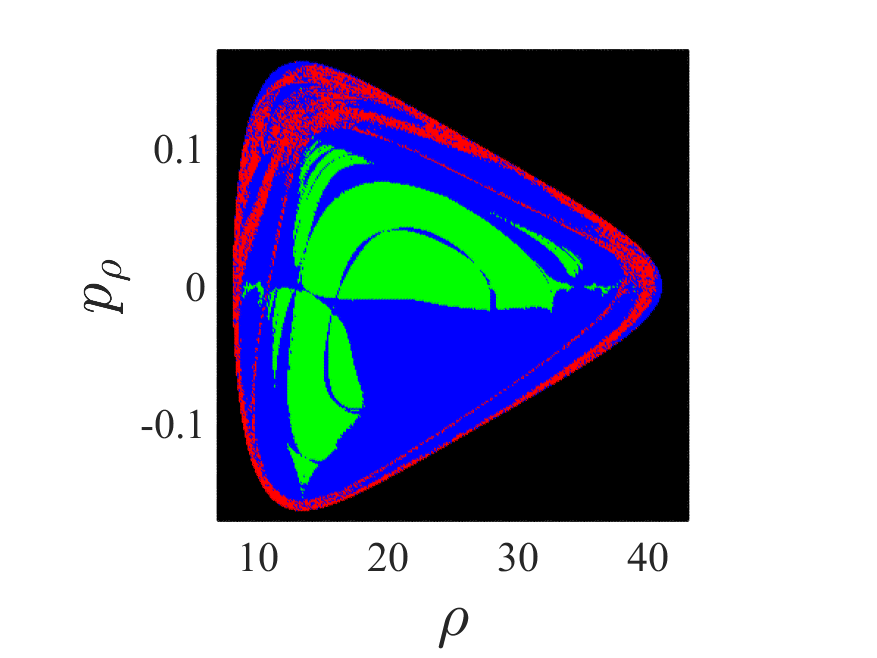}
        \caption{$a=0.2$, $D$=0 \newline $Q=4.2\times10^{-6}$, $O=5.9\times10^{-8}$}
        \label{fig:p-r_plane_spin02Q+O}
    \end{subfigure}
        \hfill
        \begin{subfigure}[b]{0.32\linewidth}
        \centering
        \includegraphics[width=2.2in]{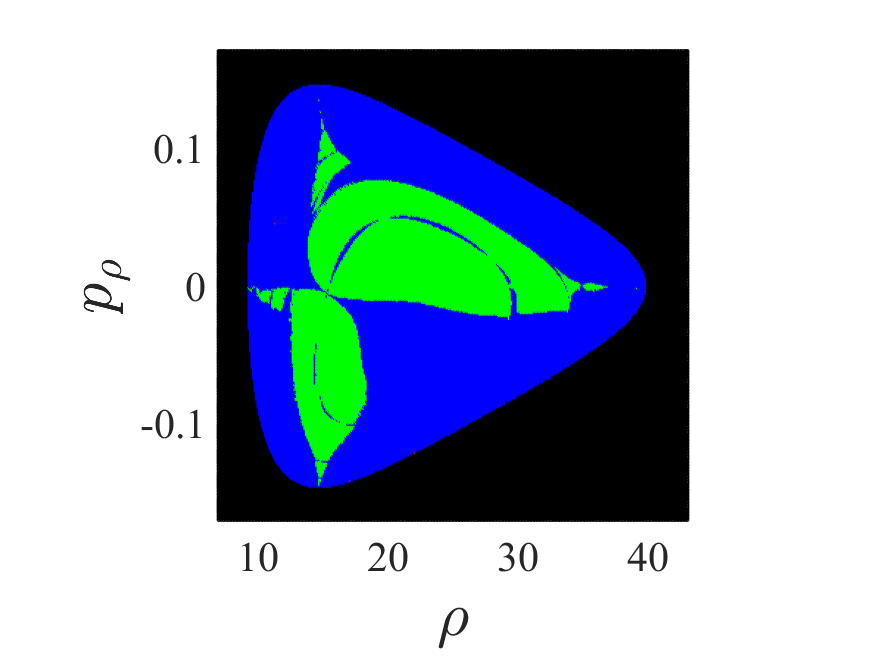}
        \caption{$a=0.5$, $D$=0 \newline $Q=4.2\times10^{-6}$, $O=5.9\times10^{-8}$}
        \label{fig:p-r_plane_spin05Q+O}
    \end{subfigure}
        \hfill
        \begin{subfigure}[b]{0.32\linewidth}
        \centering
        \includegraphics[width=2.2in]{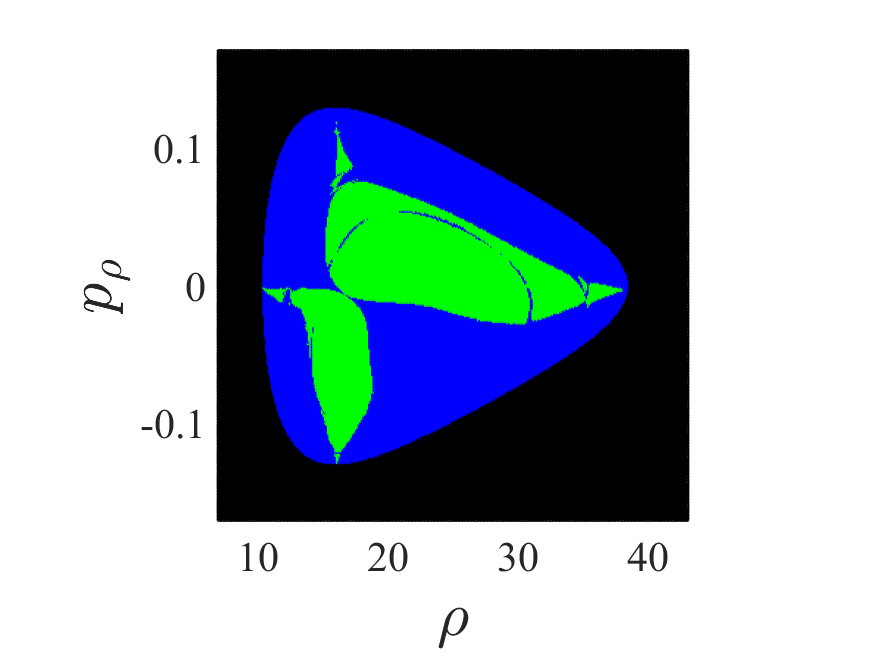}
        \caption{$a=0.8$, $D$=0 \newline $Q=4.2\times10^{-6}$, $O=5.9\times10^{-8}$ }
        \label{fig:p-r_plane_spin08Q+O}
    \end{subfigure}
        \begin{subfigure}[b]{0.32\linewidth}
        \centering
        \includegraphics[width=2.2in]{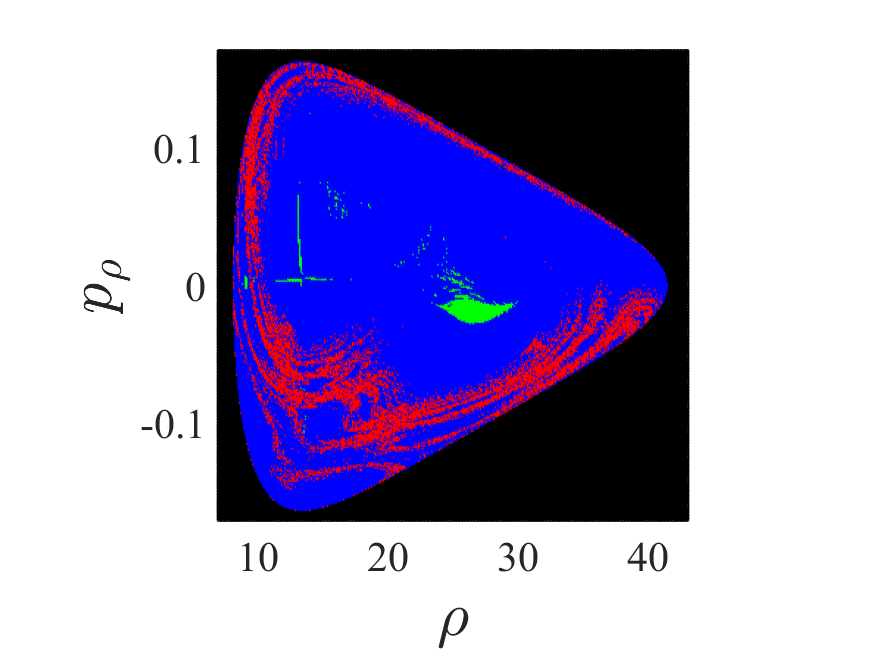}
        \caption{$a=0.2$, $O=0$ \newline $D=2\times10^{-4}$, $Q=4.3\times10^{-6}$}
        \label{fig:p-r_plane_spin02D+Q}
    \end{subfigure}
    \hfill
        \begin{subfigure}[b]{0.32\linewidth}
        \centering
        \includegraphics[width=2.2in]{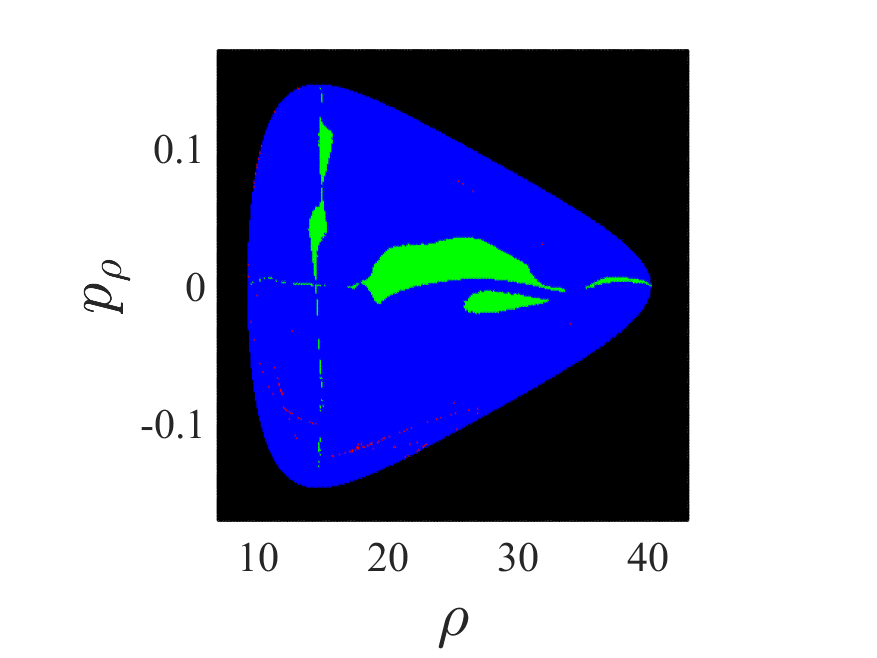}
        \caption{$a=0.5$, $O=0$ \newline $D=2\times10^{-4}$, $Q=4.3\times10^{-6}$}
        \label{fig:p-r_plane_spin05D+Q}
    \end{subfigure}
    \hfill
        \begin{subfigure}[b]{0.32\linewidth}
        \centering
        \includegraphics[width=2.2in]{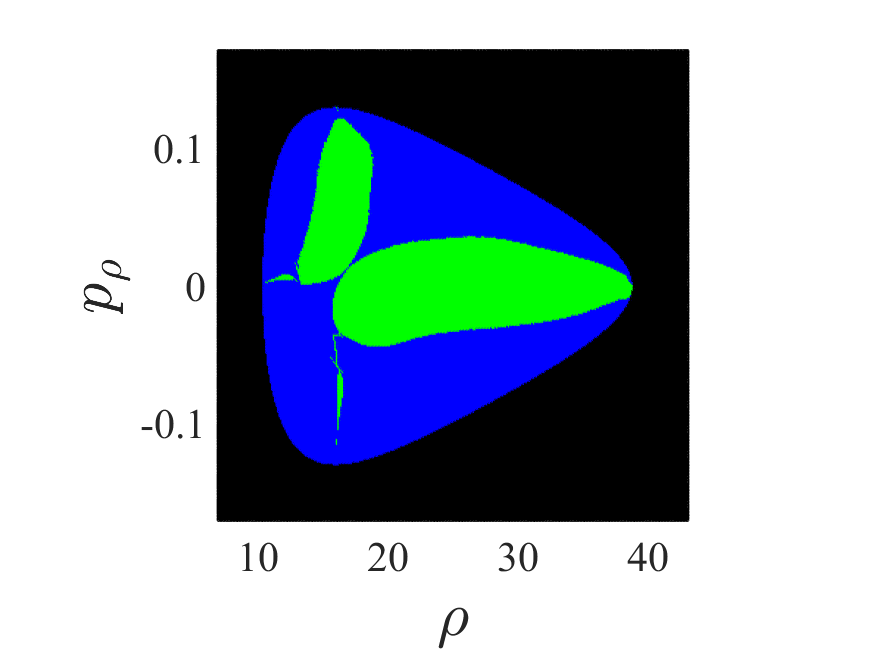}
        \caption{$a=0.8$, $O=0$ \newline $D=2\times10^{-4}$, $Q=4.3\times10^{-6}$}
        \label{fig:p-r_plane_spin08D+Q}
    \end{subfigure}
    \hfill
        \begin{subfigure}[b]{0.32\linewidth}
        \centering
        \includegraphics[width=2.2in]{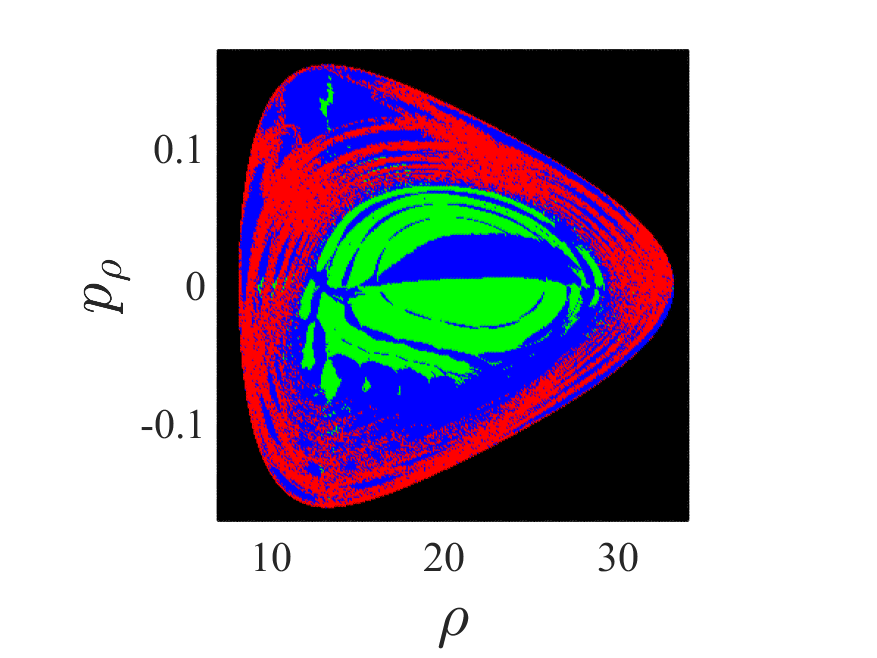}
        \caption{$a=0.2$, $Q$=0 \newline $D=2\times10^{-4}$, $O=2.4\times10^-7$}
        \label{fig:p-r_plane_spin02D+O}
    \end{subfigure}
    \hfill
        \begin{subfigure}[b]{0.32\linewidth}
        \centering
        \includegraphics[width=2.2in]{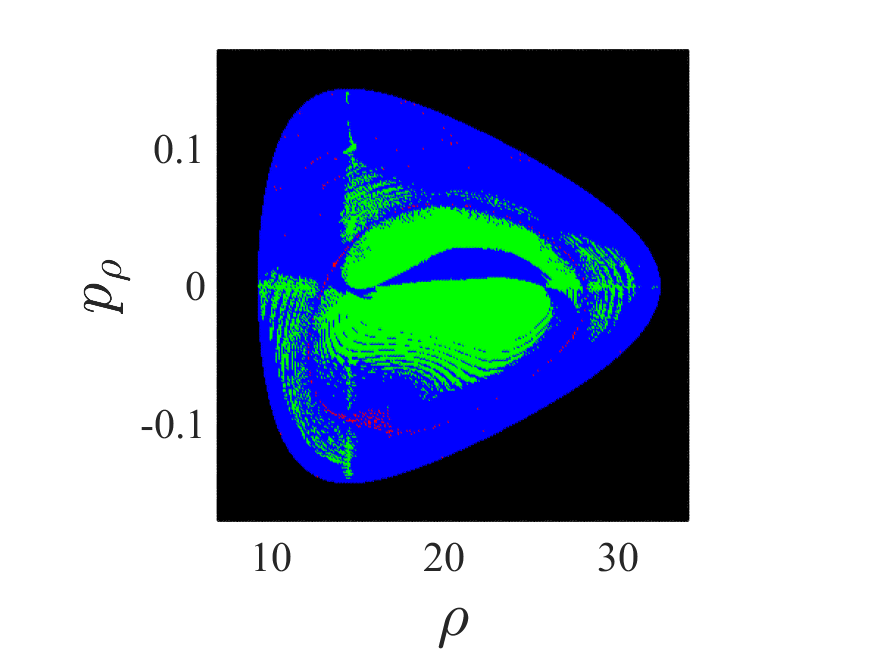}
        \caption{$a=0.5$, $Q$=0 \newline $D=2\times10^{-4}$, $O=2.4\times10^-7$}
        \label{fig:p-r_plane_spin05D+O}
    \end{subfigure}
    \hfill
        \begin{subfigure}[b]{0.32\linewidth}
        \centering
        \includegraphics[width=2.2in]{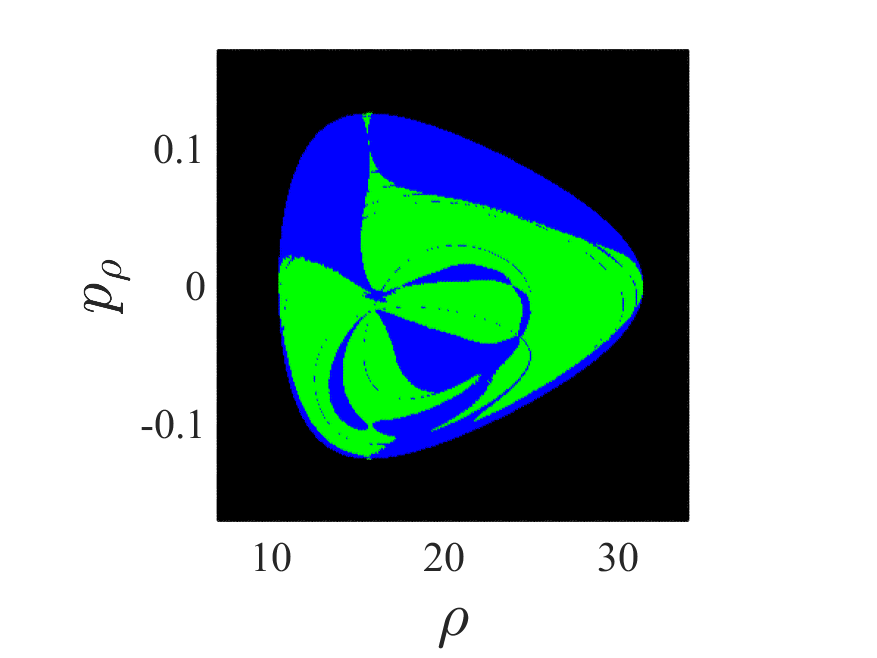}
        \caption{$a=0.8$, $Q$=0 \newline $D=2\times10^{-4}$, $O=2.4\times10^-7$}
        \label{fig:p-r_plane_spin08D+O}
    \end{subfigure}
        \hfill
        \begin{subfigure}[b]{0.32\linewidth}
        \centering
        \includegraphics[width=2.2in]{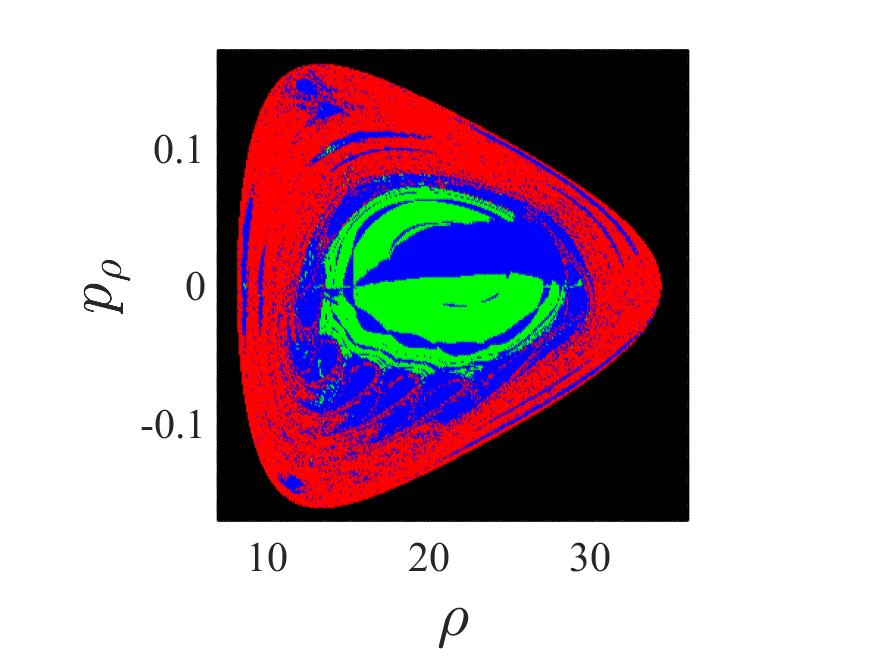}
        \caption{$a=0.2$, $D=2\times10^{-4}$ \newline $Q=1\times10^{-6}$, $O=2.2\times10^-7$}
        \label{fig:p-r_plane_spin02D+Q+O}
    \end{subfigure}
        \hfill
        \begin{subfigure}[b]{0.32\linewidth}
        \centering
        \includegraphics[width=2.2in]{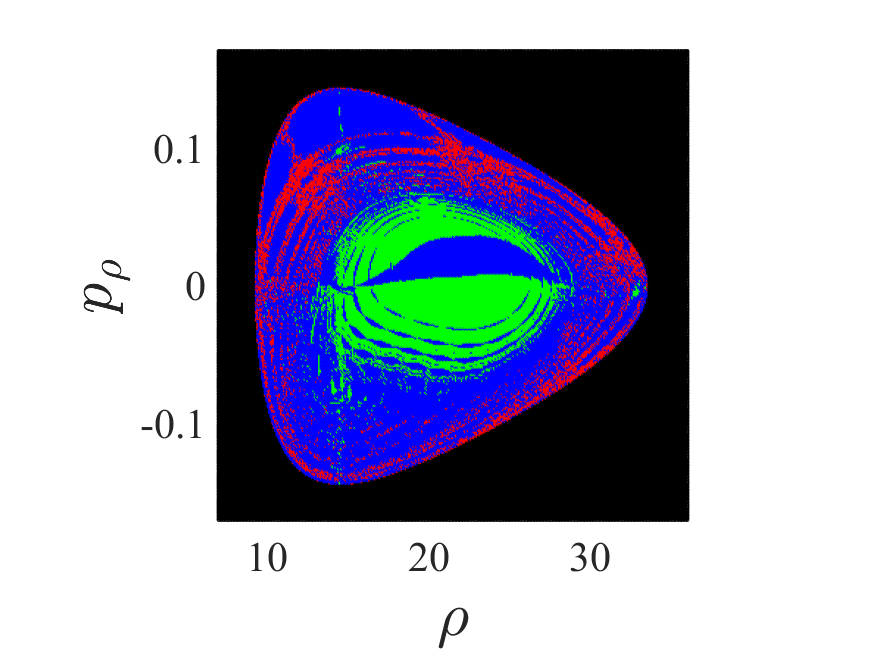}
        \caption{$a=0.5$, $D=2\times10^{-4}$ \newline $Q=1\times10^{-6}$, $O=2.2\times10^-7$}
        \label{fig:p-r_plane_spin05D+Q+O}
    \end{subfigure}
        \hfill
        \begin{subfigure}[b]{0.32\linewidth}
        \centering
        \includegraphics[width=2.2in]{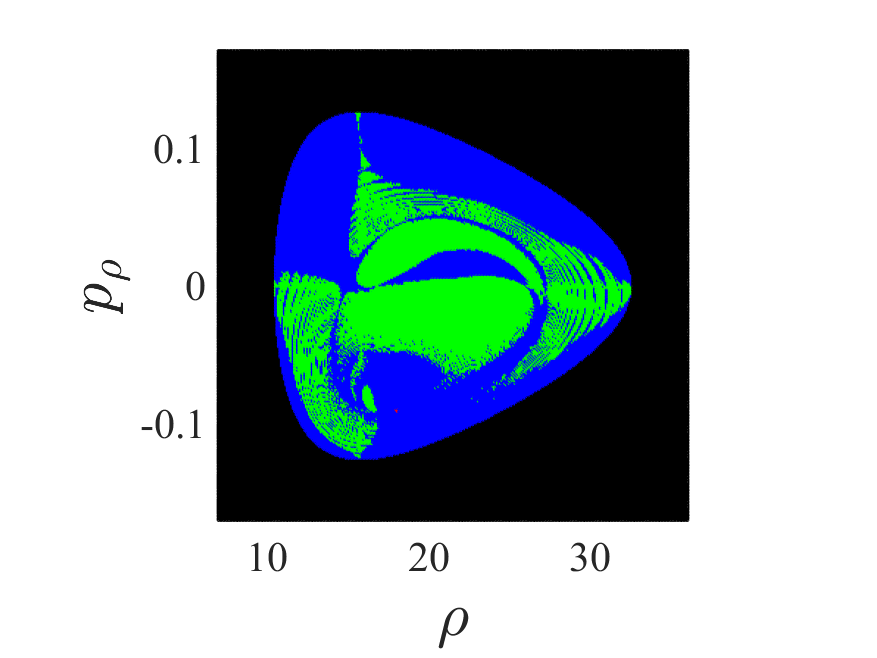}
        \caption{$a=0.8$, $D=2\times10^{-4}$ \newline $Q=1\times10^{-6}$, $O=2.2\times10^-7$}
        \label{fig:p-r_plane_spin08D+Q+O}
    \end{subfigure}
    \captionsetup{justification=raggedright,singlelinecheck=false}
    \caption{ Color-coded diagrams in the $p_{\rho}-\rho$ plane depict systems with varying spin parameters: $Q+O$ system ((\ref{fig:p-r_plane_spin02Q+O}), (\ref{fig:p-r_plane_spin05Q+O}), (\ref{fig:p-r_plane_spin08Q+O})); $Q+D$ system ((\ref{fig:p-r_plane_spin02D+Q}), (\ref{fig:p-r_plane_spin05D+Q}), (\ref{fig:p-r_plane_spin08D+Q})); $D+O$ system ((\ref{fig:p-r_plane_spin02D+O}), (\ref{fig:p-r_plane_spin05D+O}), (\ref{fig:p-r_plane_spin08D+O})) and $D+Q+O$ system ((\ref{fig:p-r_plane_spin02D+Q+O}), (\ref{fig:p-r_plane_spin05D+Q+O}), (\ref{fig:p-r_plane_spin08D+Q+O})), respectively. $L=4.2$ and $E=0.976$ are same for all the plots.}
    \label{fig:p-r_plane_CombinationMP_spin_variation}
\end{figure*}


In this investigation, our aim is to scrutinize the dynamic behavior of the system across diverse combinations of multipolar terms and spin within the \(\rho-p_{\rho}\) plane. This inquiry facilitates a visual and qualitative analysis of the system's evolution within the framework of Newtonian dynamics across different multipolar configurations. In Figure (\ref{fig:p-r_plane_OnlyMP_spin_variation}), we have investigated the impact of individual multipolar terms on spin. In figure (\ref{fig:p-r_plane_CombinationMP_spin_variation}) we have extended this analysis to examine the influence of various combinations of multipolar values. Fig.~(\ref{fig:p-r_plane_OnlyMP_spin_variation}) and Fig.~(\ref{fig:p-r_plane_CombinationMP_spin_variation}) have been generated by choosing multipolar values from Fig.~(\ref{fig:D-r_plane_SpinVariation}), Fig.~(\ref{fig:Q-r_Q_variation}) and Fig.~(\ref{fig:O-r_O_variation}) for which we get atleast one or more escaping orbits. This facilitates a comparative analysis of the system under various multipolar configurations including dipolar, quadrupolar and octupolar their combinations. For this study, we set the values of \( z = 0 \) then calculated the \( p_{z} = 0 \) value using Eq.~(\ref{eqn:HillEq_Newt}) for every pair of \( \rho \) and \( p_{\rho} \) in a 320 $\times$ 320 grid ($\rho, p_{\rho}$) to generate the plots. Table~(\ref{tab:Multipole_Value_chaos&RtoS_ratio_SystemCharacteristics}) comprehensively elucidates the system's dynamics across various multipolar values and their combinations and for several values of the spin parameter of the central compact object ($a$) representing the entire range ($0\leq a\leq 1$) in terms of the associated percentage of chaotic orbits and the ratio of regular to sticky orbits. This table serves as a concise reference for a holistic understanding of the system's behaviour under different configurations.

\begin{table*}
\begin{ruledtabular}
\begin{tabular}{cccccc}
        \(D\) & \(Q\) & \(O\) & \(a\) & Chaotic orbits $\%$ & regular orbits/sticky orbits  \\
        \hline
         0 & $5.2\times10^{-6}$ & 0 & 0.2 &  0.17 & 0.34  \\
         0 & $5.2\times10^{-6}$ & 0 & 0.5 &  0.01 & 0.28  \\
         0 & $5.2\times10^{-6}$ & 0 & 0.8 &  0 & 0.27  \\
         0 & 0 & $1.4\times10^{-7}$ & 0.2 &  0.23 & 0.63  \\
         0 & 0 & $1.4\times10^{-7}$ & 0.5 &  0 & 0.77  \\
         0 & 0 & $1.4\times10^{-7}$ & 0.8 &  0 & 0.68  \\
        $2.8\times10^{-4}$ & 0 & 0 & 0.2 & 18.41 & 0.11 \\
         $2.8\times10^{-4}$ & 0 & 0 & 0.5 & 0.01 & 0.12  \\
         $2.8\times10^{-4}$ & 0 & 0 & 0.8 &  0 & 0.11  \\
         0 & $4.2\times10^{-6}$& $0.59\times10^{-7}$& 0.2 &  12.86 & 0.50  \\
         0 & $4.2\times10^{-6}$& $0.59\times10^{-7}$& 0.5 &  0.01 & 0.50 \\
         0 & $4.2\times10^{-6}$& $0.59\times10^{-7}$& 0.8 &  0 & 0.54  \\
         $2\times10^{-4}$ & $4.3\times10^{-6}$& 0 & 0.2 &  16.53 & 0.02  \\
         $2\times10^{-4}$ & $4.3\times10^{-6}$& 0 & 0.5 &  0.20 & 0.09  \\
         $2\times10^{-4}$ & $4.3\times10^{-6}$& 0 & 0.8 &  0 & 0.53  \\
         $2\times10^{-4}$ & 0 & $2.4\times10^{-7}$ & 0.2 &  33.13 & 0.50  \\
         $2\times10^{-4}$ & 0 & $2.4\times10^{-7}$ & 0.5 &  0.51 & 0.45  \\
         $2\times10^{-4}$ & 0 & $2.4\times10^{-7}$ & 0.8 &  0 & 1.33  \\
         $2\times10^{-4}$ & $1\times10^{-6}$& $2.2\times10^{-7}$& 0.2 &  48.97 & 0.50  \\
         $2\times10^{-4}$ & $1\times10^{-6}$& $2.2\times10^{-7}$& 0.5 &  14.43 & 0.35  \\
         $2\times10^{-4}$ & $1\times10^{-6}$& $2.2\times10^{-7}$& 0.8 &  0.01 & 0.73  \\   
    \end{tabular}
    \captionsetup{justification=raggedright,singlelinecheck=false}
    \caption{Tabular representation of the different multipolar moment values, spin, the associated percentage of chaotic orbits, and the ratio of regular to sticky orbits.}
    \label{tab:Multipole_Value_chaos&RtoS_ratio_SystemCharacteristics}
\end{ruledtabular}
\end{table*}

The analysis of Fig.~(\ref{fig:p-r_plane_spin02D0O0}), Fig.~(\ref{fig:p-r_plane_spin02D0Q0}) and Fig.~(\ref{fig:p-r_plane_spin02OnlyD}) indicates that the presence of dipolar, quadrupolar, and octupolar terms singularly induces chaos in the system, particularly at low spin values ($a$ $<$ 0.5). Notably, the dipolar term exhibits the highest impact on chaos, followed by the octupolar and quadrupolar terms. Fig.~(\ref{fig:p-r_plane_spin02Q+O}), Fig.~(\ref{fig:p-r_plane_spin02D+Q}), Fig.~(\ref{fig:p-r_plane_spin02D+O}) and Fig.~(\ref{fig:p-r_plane_spin02D+Q+O}) illustrate that combinations of different multipolar values leading to amplified chaos, especially with the addition of dipolar ($D$) and octupolar ($O$) terms. The $D+Q+O$ system exhibits the highest level of chaos, followed by $D+O$, $D+Q$, and $Q+O$ systems. This suggests that the introduction of the dipolar term significantly contributes to chaos compared to the octupolar term. Here we also note that a noticeable reduction in chaotic orbits happen as the value of the spin parameter ($a$) is increased for all combinations of multipolar terms. In particular, we see that chaotic orbits in this system reduce by almost \(90\%\) or more as the value of the spin parameter ($a$) is increased from 0.2 to 0.5 and further for all cases except in the case where the dipolar, quadrupolar and the octupolar terms ($D+Q+O$) are acting simultaneously. 
This aligns with the findings of earlier work (\cite{Ying2012}), supporting the occurrence of chaos for small values of $a$. However, Interestingly we find that inclusion of all the three multipolar moments ($D+Q+O$) into the system weakens the influence of spin on the appearance of chaotic orbits, for example the reduction is only approximately \(70\%\) as the spin value increases from 0.2 to 0.5 and further as compared to  \(90\%\) in all the other cases as discussed above (refer to  Fig.~(\ref{fig:p-r_plane_spin05Q+O}), Fig.~(\ref{fig:p-r_plane_spin05D+Q}), Fig.~(\ref{fig:p-r_plane_spin05D+O}) and Fig.~(\ref{fig:p-r_plane_spin05D+Q+O})). In Fig.(\ref{fig:p-r_plane_spin08D+Q+O}), we see that some chaotic orbits are even evident at high spin value at $a=0.8$ (refer to Fig.~(\ref{fig:p-r_plane_spin08D+Q+O})). 


This trend of reduction in chaos in the system with the increase of spin for all multipolar combinations is not uniformly applicable to all kinds of orbits. Notably, we observe a systematic increase in the ratio of regular to sticky orbits (R/S) as spin values progress from low to high (a=0.2 to a=0.8) for nearly all multipolar combinations, except the $Q$-only system. In the case of the $Q$-only system, we note a steady decline in this ratio, with a significant \(21\%\) decrease observed as the spin value escalates from 0.2 to 0.8 (refer to Fig.~(\ref{fig:p-r_plane_spin02D0O0}) and Fig.~(\ref{fig:p-r_plane_spin08D0O0})). Additionally, analysis of Fig.~(\ref{fig:p-r_plane_OnlyMP_spin_variation}) and Fig.~(\ref{fig:p-r_plane_CombinationMP_spin_variation}) (Fig.~(\ref{fig:p-r_plane_spin02D0Q0}), Fig.~(\ref{fig:p-r_plane_spin05D0Q0}), Fig.~(\ref{fig:p-r_plane_spin08D0Q0}), Fig.~(\ref{fig:p-r_plane_spin02Q+O}), Fig.~(\ref{fig:p-r_plane_spin05Q+O}), Fig.~(\ref{fig:p-r_plane_spin08Q+O}), Fig.~(\ref{fig:p-r_plane_spin02D+O}), Fig.~(\ref{fig:p-r_plane_spin05D+O}), Fig.~(\ref{fig:p-r_plane_spin08D+O}), Fig.~(\ref{fig:p-r_plane_spin02D+Q+O}), Fig.~(\ref{fig:p-r_plane_spin05D+Q+O}), Fig.~(\ref{fig:p-r_plane_spin08D+Q+O})) indicates that presence of the octupolar term enhances the likelihood of regular orbits irrespective of the spin value. 

Fig.~(\ref{fig:p-r_plane_spin02D+Q}), Fig.~(\ref{fig:p-r_plane_spin05D+Q}), Fig.~(\ref{fig:p-r_plane_spin08D+Q}), Fig.~(\ref{fig:p-r_plane_spin02D+O}), Fig.~(\ref{fig:p-r_plane_spin05D+O}), Fig.~(\ref{fig:p-r_plane_spin08D+O}), Fig.~(\ref{fig:p-r_plane_spin02D+Q+O}), Fig.~(\ref{fig:p-r_plane_spin05D+Q+O}) and Fig.~(\ref{fig:p-r_plane_spin08D+Q+O}) represent the $D+Q$, $D+O$ and the $D+Q+O$ systems with increasing spin values respectively. On examining these figures, it can be concluded that the maximum variation in the R/S ratio with spin is observed in these the $D+Q$, $D+O$ and the $D+Q+O$ systems. Specifically, this ratio increases rapidly with spin for these systems with the $D+Q$ system exhibiting the greatest increase, followed by $D+O$ and $D+Q+O$ systems. This outcome implies that the inclusion of the dipolar term, when combined with others, renders the appearance of regular orbits in these systems more sensitive to spin. In contrast, the inclusion of only the dipolar term, in addition to the central monopole, results in lower sensitivity to spin and increases the likelihood of sticky orbits.

\section{Conclusion}

In this study, the dynamics of a test prticle in the presence of a central compact object along with a halo mass distribution around it in the center of a galaxy was explored through a multipolar expension of the potential due to this mass distribution. The central compact object was modeled using the pseudo-Newtonian scheme to emulate a Schwarzschild-like as well as a Kerr-like object. The motion of the test particle in this earlier described potential and its interaction with the spin of the central compact object was thoroughly explored using Poincaré section analysis and the Smaller Alignment Index (SALI). The investigation yielded several significant results which were not only in accordance with earlier findings but also reveled many new aspects through this study.

Through the analysis of the Poincaré section, we discovered that the inclusion of the dipolar and the octupolar terms intensifies the chaotic behavior in this systems containing a centrally positioned Kerr-like compact object, primarily by disrupting the reflection symmetry about the equatorial plane, as had been concluded for a simple Newtonian case (\cite{Vieira1999CoreShellGR}). A comparison between Newtonian and special relativistic formulations reveled that systems incorporating dipolar and octupolar terms tend to exhibit increased chaotic orbits in the case of Newtonian dynamics, particularly for low spin values, which contrasts with the behavior observed in systems involving other combinations of multipolar terms.

The study of this system using SALI in $Q-\rho$ plane (\ref{SubSec:Q-r}) shows a decrease in the bound-to-unbound orbit ratio with increasing spin for systems having only the quadrupolar term ($Q$-only) along with a central monopole. This behaviour is in contrast with all other combinations of multipolar terms where it is seen that the number of regular orbits increase with spin. When the dipole term is activated (i.e., in a $D+Q$ system), the regular orbits exhibit a gradual decline after reaching a maximum number as 
the scaled coordinate $\rho$ increases. This decline is slower than in the $Q$-only system. Furthermore, we have observed that the rate of decline is influenced by the value of $D$, with higher values resulting in a slower gradient. Furthermore, our analysis of the SALI plots from the $p_{\rho}-\rho$ plane (\ref{SubSec:p-r}) revealed a systematic decrease in the ratio of regular to sticky orbits as spin values increased. These findings are possibly indicative of the fact that the interaction of the quadrupolar term with the spin of the central object is most non-trivial and probably reverses the effect of spin on the dynamics of test particles, as compared to what had been usually observed by earlier authors (\cite{sankha_Nag2017}, \cite{Ying2012}, \cite{Vieira1999CoreShellGR}, \cite{Dubeibe2021Quadrupole+Octupole}). 

Based on our analysis of the plots in  the $p_{\rho}-\rho$ plane (\ref{SubSec:p-r}), generated using SALI, we have ranked different cases according to their chaotic behavior, with the $D+Q+O$ system identified as the most chaotic, followed by the $D+O$, $D+Q$, $Q+O$, $D$, $O$, and $Q$ terms. Additionally, our observations have also shown that incorporating all three multipolar moments ($D+Q+O$ system) into the system reduces the impact of spin on the emergence of chaotic orbits. Additionally, our analysis of the SALI plots revealed a systematic increase in the ratio of regular to sticky orbits as spin values increased across most systems, with the exception of the $Q$-only system, which displayed a steady decline in this ratio. 

In the O$-\rho$ plane (\ref{SubSec:O-r}), we have noted a significant rise in the chaotic to escaping orbits ratio (C/E) with increasing spin until a specific threshold, beyond which the ratio has declined. Meanwhile, regular and sticky orbits have exhibited an upward trend with spin, and the introduction of the dipolar term had marginal effects on the system's behaviour. Additionally, our analysis in subsection \ref{SubSec:p-r}, focusing on the $p_{\rho}-\rho$ plane, indicated that the presence of the octupolar term enhances the likelihood of regular orbits irrespective of the spin value.

In the $D-\rho$ plane (\ref{SubSec:D-r}), we have observed an increase in the prevalence of sticky and chaotic orbits with increasing spin. Additionally, we have identified isolated islands of bound orbits within the escaping region of escaping orbits for higher spin values ($a \geq 0.5$). Similar islands were also observed for the $O$-only system in the $O-\rho$ plane (\ref{SubSec:O-r}), but for lower spin values ($a \leq 0.5$). However, in the $p_{\rho}-\rho$ plane (\ref{SubSec:p-r}), we have found that the inclusion of the dipolar term alongside others has increased the dependence of the appearance of regular orbits on spin in these cases. Conversely, the dipolar term alone alongside the central monopole shows reduced spin sensitivity and increases the likelihood of sticky orbits.

Developing a model of galactic potential using the multipole expansion of mass distribution up to the octupole term has significant astrophysical implications. This refined model allows for a more accurate representation of non-spherical galaxies, capturing detailed structures like bars, spirals, and bulges \cite{binney2008galactic}. It enhances our understanding of stellar orbits and gas dynamics, which is crucial for studying galaxy dynamics and star formation \cite{Meiron2014ExpansionTechniquesStellarDynamical}. Furthermore, in scenarios of galaxy mergers and tidal interactions, the model aids in simulating complex dynamics, offering better comprehension of mass stripping from satellite galaxies. 
In the central region of galaxies, the gravitational potential is primarily influenced by the central supermassive black hole, causing stars to follow quasi-Keplerian orbits. These orbits are distorted by gravitational interactions with other stars, resulting in long-term orbital relaxation. Accurately simulating these processes numerically is highly challenging. By applying this model to various galaxy types and cluster environments, it helps elucidate the importance of different multipole terms, making it a valuable tool in the study of galactic and cosmological phenomena. 
The development and refinement of these models, aiming for increased precision in predicting the chaotic nature of orbits in galactic centers, can be linked to predicting nonlinearity in X-ray observations of accreting sources.

In future, we would also like to examine the variation in the dynamics of the test particle due to the presence of  prolate quadrupolar term using SALI. It is also contextual to undertake a study to explore the different tools known in current literature to estimate chaos quantitatively to understand their comparative performance for such nonlinear astrophysical  systems (\cite{Maffione2011ChaoticIndicatorComparison}, \cite{Maffione2013ChaoticIndicatorComparison}, \cite{Carpintero2014LPVIcodeAP}, \cite{Skokos2016BookChaos&Predictability}). The study of integrability conditions and the potential transition to chaotic motion for charged particles navigating under the gravitational pull of compact objects and their magnetospheres, alongside the magnetic force, has been investigated within the context of general relativity (\cite{Kovář2008ChargeParticle}, \cite{Kovář2010ChargeParticle}, 
\cite{Kopáček2014ChargeParticle}, \cite{Kopáček2015ChargeParticle}). We also plan to extend our studies to explore the chaotic motion of charged particles influenced by both, external magnetic fields and the gravitational field of the central compact objects surrounded by a multipolar halo mass distribution using the pseudo-Newtonian scheme. 
\section{Acknowledgement}
The authors express their gratitude to Dr. Soumen Roy and Roopkatha Banerjee for their invaluable computational work. Additionally, the authors would like to acknowledge the anonymous referee for their insightful comments and suggestions.

\begin{appendices}
\section{Appendix}

$\mathcal{L}=-\frac{1}{\gamma}-\Phi_g$, where $\gamma=\dfrac{1}{\sqrt{1-(\dot{\rho}^2+\rho^2\dot{\phi}^2+\dot{z}^2)}}$ \\

So, $\mathcal{L}=-{\sqrt{1-(\dot{\rho}^2+\rho^2\dot{\phi}^2+\dot{z}^2)}}-\Phi_g$
Now 
\begin{align*}
    H=& p\dot{q}-\mathcal{L} \\
    =& {p_{\rho}}{\dot{\rho}}+{p_{\phi}}{\dot{\phi}}+{p_z}{\dot{z}}+{\sqrt{1-(\dot{\rho}^2+\rho^2\dot{\phi}^2+\dot{z}^2)}}+\Phi_g \\
    =& \frac{\dot{\rho}^2+{\rho}^2\dot{\phi}^2+{\dot{z}}^2}{\sqrt{1-(\dot{\rho}^2+\rho^2\dot{\phi}^2+\dot{z}^2)}} + \\ & {\sqrt{1-(\dot{\rho}^2+\rho^2\dot{\phi}^2+\dot{z}^2)}}+\Phi_g \\
    =& \frac{\dot{\rho}^2+{\rho}^2\dot{\phi}^2+{\dot{z}}^2+1-(\dot{\rho}^2+\rho^2\dot{\phi}^2+\dot{z}^2)}{\sqrt{1-(\dot{\rho}^2+\rho^2\dot{\phi}^2+\dot{z}^2)}} +\Phi_g \\
    =& \frac{1-(\dot{\rho}^2+\rho^2\dot{\phi}^2+\dot{z}^2)}{\sqrt{1-(\dot{\rho}^2+\rho^2\dot{\phi}^2+\dot{z}^2)}} +\Phi_g \\
    =& {\sqrt{1-(\dot{\rho}^2+\rho^2\dot{\phi}^2+\dot{z}^2)}} +\Phi_g \\
  \implies H=& \gamma + \Phi_g
\end{align*}

\end{appendices}

\nocite{*}
\bibliographystyle{aps}
\bibliography{ref}
\end{document}